\documentclass[12pt]{article}
\usepackage{amsmath,amssymb,epsfig}
\usepackage{graphicx,floatflt,subfigure}
\usepackage{cite}

\makeatletter

\def\half{{\frac{1}{2}}}

\def\unit{{1\kern-.65ex {\rm l}}}
\def\1{{1\kern-.65ex {\rm l}}}

\newcount\hour \newcount\minute
\hour=\time \divide \hour by 60
\minute=\time
\def\now{%
\ifnum \hour<13
  \ifnum \hour=0 \advance \hour by 12 \number\hour:\else \number\hour:\fi%
     \ifnum \minute<10 0\fi%
     \number\minute%
\ A.M.%
\else \advance \hour by -12 \number\hour:%
  \ifnum \minute<10 0\fi%
  \number\minute%
  \ P.M.%
\fi%
}

\makeatother

\begin{document}

\baselineskip=18pt  
\numberwithin{equation}{section}  
\allowdisplaybreaks  



%
%


\thispagestyle{empty}

\vspace*{-2cm}
\begin{flushright}
{\tt EFI-10-13}\\
{\tt NSF-KITP-10-073}\\
{\tt PI-strings-186}
\end{flushright}

\vspace*{0.8cm}
\begin{center}
 {\LARGE A Note on $G$-Fluxes for F-theory Model Building\\}
 \vspace*{1.5cm}
 Joseph Marsano$^1$, Natalia Saulina$^2$, and Sakura Sch\"afer-Nameki$^3$\\
 \vspace*{1.0cm}
{\it
$^1$ Enrico Fermi Institute, University of Chicago\\ 5640 S Ellis Ave, Chicago, IL 60637, USA\\[1ex]
$^2$ Perimeter Institute for Theoretical Physics\\ 31 Caroline St N., Waterloo, Ontario N2L 2Y5, Canada \\[1ex]
$^3$ Kavli Institute for Theoretical Physics \\ University of California, Santa Barbara, CA 93106, USA\\[1ex]
} \vspace*{0.4cm}

{\tt marsano,  saulina, ss299  theory.caltech.edu}
\end{center}
\vspace*{.5cm}

\noindent
We propose a description of $G$-fluxes that induce chirality in 4-dimensional F-theory GUT spectra that is intrinsic to F-theory and does not rely on Heterotic/F-theory duality.  Using this, we describe how to globally extend fluxes that have been constructed in a semi-local setting and obtain an F-theoretic formula for computing the chiral spectrum that they induce.  Chirality computations agree with those from the semi-local Higgs bundle analysis for matter fields that are charged under the GUT-group, and hence with the standard Heterotic formulae where applicable.  Finally, the relation of $G$-fluxes to $SU(5)_{\perp}$ bundles on the F-theory 4-fold is discussed and used to motivate a quantization rule that is consistent both with the Higgs bundle one as well as the Heterotic one when a Heterotic dual exists.




\newpage
\setcounter{page}{1} 



\tableofcontents


\section{Introduction and Summary}

While F-theory has proven to be fertile ground for engineering supersymmetric GUTs \cite{Donagi:2008ca,Beasley:2008dc,Beasley:2008kw,Donagi:2008kj}, much of our understanding of the basic tools for model building relies on the duality to Heterotic strings \cite{Morrison:1996na,Morrison:1996pp,Friedman:1997yq,Curio:1998bva,Hayashi:2008ba}.  This is made possible because the degrees of freedom that transform nontrivially under the GUT group localize near a 4-cycle, $S_2$, where the F-theory compactification develops a K3 type singularity.  GUT-charged degrees of freedom do not explore geometry outside of this region, which is locally K3-fibered, so they cannot distinguish between the actual geometry in which they live and one that is globally K3-fibered
{\footnote{The limitations of this approach are obvious; any physics involving degrees of freedom that can propagate away from $S_{\rm GUT}$ cannot be inferred from Heterotic.  The most notable example of this is the effect of $U(1)_Y$ flux, which can lift the $U(1)_Y$ boson through coupling to "closed string" axions \cite{Beasley:2008kw,Donagi:2008kj}.  Whether the $U(1)_Y$ boson is actually lifted depends on the spectrum of "closed string" axions which, in turn, depends on details of the full compactification apart from the local K3 geometry.}}.  Geometries that are globally K3-fibered, though, admit a dual Heterotic description that can be used to learn many things about the GUT-charged sector in general.  This approach has shed light on a number of subtle details of F-theory compactifications with ADE singularities that would otherwise have been quite obscure \cite{Hayashi:2008ba,Hayashi:2009ge}.

Nowhere have results from Heterotic strings been a more important "crutch" than in the construction of $G$-fluxes.  When the $F$-theory compactification is globally K3-fibered, there is a proposed map that relates certain $G$-fluxes to twisted spectral bundles on the Heterotic side \cite{Curio:1998bva}.  Unlike GUT-charged fields, though, $G$-fluxes permeate the entire compactification and are therefore sensitive to global details of the geometry away from $S_2$.
To this point, compact model building \cite{Andreas:2009uf,Marsano:2009ym,Marsano:2009gv,Blumenhagen:2009yv,Marsano:2009wr,Grimm:2009yu,Cvetic:2010rq,Chen:2010tp,Chen:2010ts} has essentially assumed that any "local" flux, whose behavior is specified only near $S_2$ using Heterotic duality, can be globally extended even when the full 4-fold itself does not admit a Heterotic dual description.

\subsection{$G$-Flux and sub-bundles of $E_8$}

This begs for a description of the $G$-fluxes that we need for model building in a manner that is intrinsic to F-theory.  In this note, we make a simple proposal for how this can be done.  Our construction is motivated by the fact that, near $S_2$ where the geometry looks locally K3-fibered, we seek $G$-fluxes that take the approximate form \cite{Donagi:2008ca}
\begin{equation}G\sim \omega_i\wedge F_i \,.
\end{equation}
Here, the $\omega_i$ are $(1,1)$-forms that are dual, in the (local) K3 fiber, to 2-cycles that degenerate along various curves inside $S_2$.  Each $\omega_i$ essentially corresponds to a (local) $U(1)$ gauge field while the corresponding $F_i$ is then interpreted as a suitable "flux" of that gauge field.  When we engineer $SU(5)$ GUT's, for instance, the $\omega_i$ transform in the fundamental representation of $SU(5)_{\perp},$ the commutant of $SU(5)_{\rm GUT}$ inside $E_8$.  In general, these elements, and consequently the $\omega_i$, are mixed by monodromies whose action is via the Weyl group of $SU(5)_{\perp}$.  What the $\omega_i$ seem to specify, then, is an $SU(5)_{\perp}$ bundle on the full 4-fold $Z_4$.

Following Friedman, Morgan, and Witten \cite{Friedman:1997yq}, we can try to construct such a bundle via the spectral cover construction.  To avoid confusion with the Heterotic spectral cover, we will refer to the F-theory object as the "spectral divisor" ${\cal{C}}$ of the F-theory 4-fold $Z_4$.  With this mechanism of specifying the $\omega_i$'s, we expect that adding the flux data, $F_i$, corresponds to "twisting" the $SU(5)_{\perp}$ bundle.  This "twist" corresponds to specifying a divisor (or equivalently a $(1,1)$-form) inside the "spectral divisor" which, via the embedding $\iota:{\cal{C}}\rightarrow Z_4$, determines a 4-cycle inside $Z_4$ and hence a $(2,2)$-form that we would like to identify{\footnote{Note that a surface inside ${\cal{C}}$ will be distinguished from a surface in the same homology class in $Z_4$ but that is not contained in ${\cal{C}}$.  This is because ${\cal{C}}$ meets the locus of $SO(10)$ singularities, so what we really mean by ${\cal{C}}$ will be a proper transform under the resolution of these singularities and similar for any surface contained therein.}} with $G.$

Of course, this general picture is only heuristic at best.  For starters, FMW \cite{Friedman:1997yq} tell us how to construct bundles on \emph{smooth} elliptically fibered manifolds while 4-folds $Z_4$ used for GUT model building are singular in general.  In fact, the spectral divisor ${\cal{C}}$ that we introduce will necessarily intersect the singular locus, meaning that a proper definition of it requires some knowledge about how the resolution takes place.  This can complicate computations involving any $G$-fluxes that are defined in this way because, in principle, we have to honestly resolve $Z_4$ before calculating anything.  Fortunately, relevant details of the resolution can all be described in the locally K3-fibered region near $S_2$ and hence are in some sense "universal".  This will enable us to sharpen the idea by starting with the standard story in a globally K3-fibered geometry with a Heterotic dual.

\subsection{$G$-Fluxes when a Heterotic dual exists}

To make the general idea more precise, we start by considering F-theory 4-folds $Z_4$ that are globally K3-fibered, a setting in which we can take advantage of the connection to Heterotic strings.  The geometry of the dual Heterotic compactification emerges from the F-theory one in a stable degeneration limit of the K3 fibers \cite{Morrison:1996na,Morrison:1996pp}, in which the K3 essentially splits into a pair of $dP_9$'s glued together along a common elliptic curve.  Restricting fiberwise to the elliptic curve, we find the elliptically fibered Heterotic Calabi-Yau 3-fold, ${\cal{Z}}_H$.

In addition to ${\cal{Z}}_H$, which sits in the "middle" of the K3 fibers, the F-theory geometry also contains a $dP_9$ fibration on each side of ${\cal{Z}}_H$.
As described in several places \cite{Curio:1998bva,Donagi:2008ca,Hayashi:2008ba}, the data of each $dP_9$ fibration specifies an $E_8$ bundle on ${\cal{Z}}_H$ for one of the two Heterotic $E_8$ gauge groups.  When the $dP_9$ fibration exhibits an ADE singularity, the fibration instead specifies a subbundle which breaks $E_8$ to the subgroup specified by the singularity type.  In the case of $SU(5)_{\perp}$ bundles (and other $SU(n)$ bundles), which are the primary focus of this note, there is a simple way to connect the $dP_9$ construction of Heterotic bundles to those obtained by another technique, namely that of the spectral cover \cite{Curio:1998bva,Donagi:2008ca,Hayashi:2008ba}.  This is reviewed in more detail in the main text.  For now, we just note that the basic technique, starting with a $dP_9$ fibration with an $SU(5)_{\rm GUT}$ singularity, is to identify a union of five exceptional lines in the $dP_9$ that correspond to a fundamental of $SU(5)_{\perp}$ under the standard identification of $E_8$ roots with exceptional lines that do not intersect $e_9$.  Each of these exceptional lines intersects the anticanonical class of $dP_9$, which is identified with the elliptic fiber of ${\cal{Z}}_H$, exactly once.  Fibering this intersection over the base $S_2$, we obtain the Heterotic spectral cover, ${\cal{C}}_H$, corresponding to the $SU(5)_{\perp}$ bundle specified by the $dP_9$ fibration \cite{Curio:1998bva,Donagi:2008ca,Hayashi:2008ba}.

From this description, it is clear that the union of five exceptional lines, fibered over $S_2$, is precisely the object that should correspond to the "spectral divisor" ${\cal{C}}$.  It is also clear that to reproduce a twist of the Heterotic spectral bundle, which corresponds to a $(1,1)$-form $\gamma_H$ on ${\cal{C}}_H$, we should introduce a $(1,1)$-form $\gamma$ on ${\cal{C}}$ that restricts to $\gamma_H$ on ${\cal{C}}_H$.  This approach to $G$-fluxes in K3-fibered geometries is of course not new.  The authors of \cite{Curio:1998bva} refer to this union of lines as part of a "cylinder", $R$, and defines a projection map $p_R$ that sends $R\rightarrow {\cal{C}}_H$. Then, the $G$-flux is usually obtained from $\gamma_H$  via $i_*p_R^*\gamma_H$ \cite{Curio:1998bva,Donagi:2008ca,Hayashi:2008ba}.  For us, ${\cal{C}}$ is essentially (part of) the cylinder. However,
our spectral flux $\gamma$ does not appear to be the same  as $p_R^*\gamma_H$. 
In general, there are several different $\gamma$'s on ${\cal{C}}$ that all restrict to $\gamma_H$ on ${\cal{C}}_H$. \footnote{While this distinction does not matter for Heterotic computations, it is crucially important for calculations of spectra to work out properly when we do them intrinsically in F-theory, as they are performed in the local geometry near the singular locus rather than ${\cal{Z}}_H$.}  Ultimately, we will choose one by imposing an analog of the "traceless" constraint, motivated largely by the desire to connect to $SU(5)_{\perp}$ bundles on the full 4-fold $Z_4${\footnote{The difference between the $\gamma$ that we obtain this way and our understanding of $\iota_*p_R^*\gamma_H$ is only visible at the $SU(5)_{\rm GUT}$ singular locus.  One may think of this prescription as a proper realization of $\iota_*p_R^*\gamma_H$ that (homologically) accounts for behavior at the singularity.  After all, if we do not worry about the singular locus, the object $\iota_*p_R^*\gamma_H$ naively seems like it should be traceless.}}.   This constraint is what will ultimately make it possible to connect chirality computations in F-theory with their Heterotic counterparts, thereby guaranteeing anomaly-free spectra{\footnote{Obtaining a spectrum that is free of gauge anomalies is really the only constraint that should be imposed.  $G$-fluxes that are "traceless" will map to twists $\gamma_H$ on the Heterotic side that are also "traceless" in that context, where it is known explicitly that "traceless" implies anomaly-free \cite{Donagi:2004ia}.}}.


Given the proposal for realizing $G$-fluxes in this setting, we then turn to an intrinsically F-theoretic approach to the counting of chiral matter.  As described in \cite{Hayashi:2008ba}, this should be done by integrating the $G$-flux over a suitable "matter surface" $\hat{\Sigma}_R$, obtained by starting with a matter curve $\Sigma_R$ inside $S_2$ where the singularity type of the K3 fiber enhances and combining it with the (resolved) 2-cycle whose degeneration led to the enhancement
\begin{equation}\chi(R) = \int_{\hat{\Sigma}_R}\,G \,.
\end{equation}

A computation of this sort naively seems to depend strongly on details of the resolution, though, which leads to a small puzzle.  Heterotic/F-theory duality tells us that there are general formulae for chiralities that follow from certain "universal" fluxes that can be expressed in terms of data of the singular F-theory 4-fold.  That this happens is not unrelated to the fact that the Heterotic computation can be formulated directly in the F-theory geometry, in a sense, but the place where ${\cal{Z}}_H$ sits is "very far" from the singular loci.
Any F-theoretic description of $G$-fluxes should explain what a computation in the "middle" of the K3, where ${\cal{Z}}_H$ resides, has to do with the number of light fields at the singularities, which lie deep inside the $dP_9$'s.



 Fortunately, it is easy to describe how the fluxes that we construct should behave after the resolution in this setting using simple facts about the geometry of K3 or, more specifically, $dP_9$.  Using only local intersection data near the singular locus, we are able to formulate the problem of counting chiral matter in a very familiar way.  As we said, the $G$-flux is described as a $(1,1)$-form $\gamma$ inside ${\cal{C}}$.  For each charged representation $R$, we also identify a distinguished surface $\tilde{\Sigma}_R$ inside ${\cal{C}}$ that we call the "dual matter surface" in order to make clear that it is distinct from the true "matter surface" $\hat{\Sigma}_R$ described above.  The chiral spectrum is then determined by identifying a distinguished curve $C_{\tilde{\Sigma}_R}$ in the "dual matter surface" with certain properties and integrating $\gamma$ over it.  In essence, then, we get an expression of the form
 \begin{equation}\chi(R) = \int_{\hat{\Sigma}_R} G  = \int_{C_{\tilde{\Sigma}_R}}\gamma \,.
 \end{equation}
 In general, there will only be one distinguished curve of the right type $C_{\tilde{\Sigma}_R}\in\tilde{\Sigma}_R$ but when the geometry is $dP_9$-fibered, there are in fact two.  One of them lives inside ${\cal{Z}}_H$, where it is the object typically referred to as the matter curve in the Heterotic context.  The second, however, lives above the singular locus $S_2$, which is the more natural place to perform such a computation from an F-theory perspective.  It is easy to see on quite general grounds that integrating $\gamma$ over either yields an equivalent result when $\gamma$ satisfies a "traceless" condition.

\subsection{Generalization and Connection to $SU(5)_{\perp}$ Bundles}

Once we have moved the computation of chiral spectra from ${\cal{Z}}_H$ to the region near the singular locus in this way, it can be exported to more general F-theory 4-folds.  We describe how ${\cal{C}}$ should be defined in this case and then turn to a determination of the chiral spectrum induced by $G$-fluxes constructed via ${\cal{C}}$.  In deriving the corresponding results in the Heterotic setting, we were careful to make use only of local intersection data of the resolved geometry near the singular locus, which should be universal for any F-theory 4-fold with a local ADE singularity over $S_2$.  It is this that allows us to extend our prescription to more general F-theory 4-folds.

As a first check, we verify that, in the case of "universal" fluxes, the results are in complete agreement with the naive application of Heterotic formulae.  After this, though, we turn to semi-local models of the type described in \cite{Marsano:2009gv,Marsano:2009wr} which cannot be embedded into 4-folds with a Heterotic dual.  There, we describe the connection of the spectral divisor ${\cal{C}}$ to the "local" spectral cover ${\cal{C}}_{loc}$ that describes the Higgs bundle of the 8-dimensional gauge theory on the $SU(5)_{\rm GUT}$ locus \cite{Donagi:2009ra}.  Using ${\cal{C}}$, we show how to extend some of the fluxes of the semi-local models of \cite{Marsano:2009wr} to global $G$-fluxes on a particular type of Calabi-Yau 4-fold and demonstrate that our chirality formulae agree with those obtained from a Higgs bundle analysis \cite{Marsano:2009wr} for matter fields charged under the GUT gauge group.

Finally, with this explicit construction of $G$-fluxes as a divisor inside ${\cal{C}}$, we can then return to our original motivation, which was to relate them to $SU(5)_{\perp}$ bundles $V$ on the Calabi-Yau 4-fold, $Z_4$.  Here, we do not have much to say at the moment other than the following simple observation.  If we are cavalier about subtleties associated to the singular locus, then integral quantization of $V$ implies
a quantization rule for the $G$-fluxes that is consistent with Heterotic duality.  Note that $G$-flux quantization in M-theory compactified on a Calabi-Yau 4-fold, and hence F-theory in the appropriate limit, is typically expressed in terms of $c_2(Z_4)$ \cite{Witten:1996md}.  When $Z_4$ is singular, though, we do not know of a unique way to define the Chern classes.  To proceed, one needs a "physical" notion for what the correct definition of $c_2(Z_4)$ should be and we believe that the connection to $SU(5)_{\perp}$ bundles provides a good candidate for this.  That we obtain consistency with the Heterotic quantization rules in this way seems promising.

\subsection{Spectral Divisor and Monodromy Structure}

Finally, let us make a few comments about the relation of ${\cal{C}}$ to monodromy groups that are important for understanding key aspects of the phenomenology of F-theory models.  In general, the data of the local K3 fibration near $S_2$ is equivalent to that of a (meromorphic) Higgs bundle of an 8-dimensional  $E_8$ gauge theory on $\mathbb{R}^{3,1}\times S_2$ \cite{Donagi:2008ca}.  Encoded in this data is a monodromy group that mixes the Abelian factors which survive the breaking $E_8\rightarrow SU(5)_{\rm GUT}\times U(1)^4$ as one moves throughout $S_2$.  This structure is crucial for phenomenology because the surviving $U(1)$'s, if there are any, provide an important means of controlling the superpotential of the low energy effective theory.

Geometrically, the $U(1)$ factors here are associated to 2-cycles that yield harmonic $(1,1)$-forms that in principle correspond to massless gauge bosons.  In general, these 2-cycles will undergo a series of monodromies in the 4-fold $Z_4$.  In an interesting recent paper\cite{Hayashi:2010zp}, it was pointed out that the Higgs bundle only captures a subset of these monodromies, in general, as it misses additional monodromies of two different kinds.  The first kind involves moving around a path that takes one outside of the local geometry near $S_2$.  The second, however, involves mixing with new 2-cycles that we don't normally associate with the local K3 singularity.  This happens as we move inside $S_2$ around points where the Higgs bundle, which is generically meromorphic, exhibits a pole.

To see why the Higgs bundle picture doesn't see all of the monodromies{\footnote{We are very grateful to T.~Watari for explaining this to us at the YITP Workshop "Branes, Strings, and Black Holes" in October, 2009.}}, let us first consider the case in which a global Heterotic dual exists.  There, the spectral cover, which is a divisor inside the elliptically fibered 3-fold ${\cal{Z}}_H$, specifies the bundle data for $SU(n)$ bundles.  Locally, each sheet corresponds roughly to an eignevalue of an $SU(n)$ adjoint matrix and the interconnectedness of the sheets describes the monodromy structure.  The corresponding Higgs bundle is specified only by the local geometry of the spectral cover near the zero section of the elliptic fibration{\footnote{Typically, this is constructed as a noncompact cover of $S_2$ in the total space of the canonical bundle, which captures the local geometry near $S_2$ inside ${\cal{Z}}_H$.}}.  If we focus only on how ${\cal{C}}_H$ behaves near the zero section, it may appear to factor into multiple components despite the fact that those components might be connected inside the full ${\cal{C}}_H$.  In this case, the Higgs bundle will exhibit a reduced monodromy group, reflecting the fact that not all of the sheets appear to be connected.  That different components actually become connected when we consider the full ${\cal{C}}_H$ reflects the fact that the monodromy group is in fact larger in the full compactification.

As we will see, the data of ${\cal{C}}_H$ specifies that of the spectral divisor, ${\cal{C}}$, in these models so we can use factorization properties of ${\cal{C}}$ as a diagnostic for determining the monodromy structure in place of ${\cal{C}}_H$.  Because we can define ${\cal{C}}$ more generally, we can hope that ${\cal{C}}$ provides a nice tool for studying the global monodromy group when $Z_4$ is not globally K3-fibered.  We do not have a proof stating that reduction of the monodromy group happens if and only if ${\cal{C}}$ globally factors, but we can make one observation right away.  If ${\cal{C}}$ does globally factor, the monodromy group is definitely decreased.  This can be seen because ${\cal{C}}$ is a union of exceptional lines $\ell^{(i)}$ (fibered over $S_2$), which have distinct intersections with the local roots, $C^{(a)}$, from which the Abelian factors of interest arise.  If ${\cal{C}}$ factors, its components provide distinct divisors, say ${\cal{C}}^{(1)}$ and ${\cal{C}}^{(2)}$, whose intersections with the roots $C^{(a)}$ we can compute.  Because ${\cal{C}}^{(1)}$ and ${\cal{C}}^{(2)}$ describe different sets of $\ell^{(i)}$, their intersections with $C^{(a)}$ will in general be different.  Pairs of $C^{(a)}$ that differ in their intersections with ${\cal{C}}^{(1)}$ or ${\cal{C}}^{(2)}$ are distinct in the homology of $Z_4$ so cannot be related by monodromy.

\subsection{Future Directions}

The description of $G$-fluxes in this note opens up several new questions.  First, when ${\cal{C}}$ factors, one expects not only a reduced monodromy group and new $U(1)$ gauge symmetries but also $U(1)$-charged fields that are GUT singlets.  We make some preliminary comments about how the spectral divisor can be used to study these but a more thorough analysis that derives a counting formula would be useful.  Further, a description of precisely how the $U(1)$ gauge bosons are lifted when the spectrum is anomalous is not obvious.  As in the case of $U(1)_Y$ flux\cite{Beasley:2008kw,Donagi:2008kj}, one should be able to see explicitly that conditions for a spectrum free of $U(1)$ anomalies are necessary to obstruct the mechanism by which the $U(1)$ gauge bosons are lifted.

In this note, we do not address the D3-brane tadpole induced by $G$-fluxes.  There is some evidence \cite{Blumenhagen:2009yv} that the tadpole from the fluxes that are studied in this note is determined entirely by the behavior of the "local flux" in the local geometry near $S_2$, at least when the GUT-divisor in $B_3$ does not intersect another divisor on which non-Abelian gauge fields localize.  On one hand, this seems strange because the D3-brane tadpole involves an integration over the entire four-fold.  On the other hand, the fluxes that we study are "localized" in the sense that, homologically, they only depend on classes that arise when the $SU(5)_{\rm GUT}$ singularity is resolved.  One can guess formulae that reproduce the results that one expects from naive application of Heterotic computations but it is not clear why such formulae are correct or where they come from.  Pursuing this further would be very useful.

It would also be interesting to study the role that these fluxes may play, if any, in the structure of superpotential terms induced by D3-brane instantons{\footnote{For a recent review of D3-brane instantons in type II theories, see \cite{Blumenhagen:2009qh}.}}.  Worldvolume fluxes played a crucial role in early studies of instantons for ultra-local F-theory phenomenology \cite{Heckman:2008es,Marsano:2008py} but proper studies of both the fluxes and the instantons themselves in a compact setting were lacking.  Since then, several studies of D3-brane instantons in F-theory have been undertaken \cite{Cvetic:2009ah,Blumenhagen:2010ja,Cvetic:2010rq} and, very recently, a framework for studying them away from the weak coupling limit, in certain favorable geometric situations, has appeared \cite{Donagi:2010pd}.  It would be interesting to explore how the fluxes described here might enter that story and be used to induce phenomenologically interesting couplings.

Finally, the connection to $SU(5)_{\perp}$ bundles should be explored with more mathematical rigor.  Our goals in the present note were largely to use intuition from this connection to develop a framework for practical computations.  In doing so, we manage to avoid a truly careful treatment of how the singularity and its resolution affect the $SU(5)_{\perp}$ bundle that we presume to be defined by ${\cal{C}}$.

\textbf{Note added for v2:} After we submitted this paper to the arXiv but before it was announced, the paper \cite{Grimm:2010ez} appeared which describes many important aspects of $U(1)$ symmetries in F-theory GUT models.  Among the results of \cite{Grimm:2010ez} is a method for ensuring the existence of $U(1)$'s that couple to charged fields on matter curves.  That method seems equivalent to the criterion we mention that the spectral divisor splits into multiple components.  

\subsection{Outline}

The outline of this note is as follows.  We first review some aspects of Heterotic/F-theory duality in section 2, including the spectral cover and del Pezzo constructions of $SU(5)_{\perp}$ bundles on the Heterotic side.  Next, we describe our proposal for constructing $G$-fluxes in section 3.  There, we also derive general formulae for determining the spectrum of chiral matter and elaborate on the proposed quantization rule, which is shown to be consistent with the Heterotic one.  In section 4, we apply our understanding of $G$-fluxes to globally extend the fluxes of a class of semi-local models \cite{Marsano:2009wr} on a particular type of Calabi-Yau 4-fold.  We find complete agreement between the F-theory chirality formulae and those obtained from a Higgs bundle analysis for matter charged under the GUT group and make some comments about how the spectral divisor may be useful for studying charged singlet fields as well.


\section{$G$-Fluxes when there is a Heterotic Dual}

In this section, we recall two methods for constructing $SU(5)_{\perp}$ bundles on elliptically fibered Calabi-Yau 3-folds ${\cal{Z}}_H$, the relation between them, and some basic aspects of the Heterotic/F-theory duality map.  We include this discussion, which is a review of known results, in order to help motivate our proposal in section 3.

\subsection{$SU(5)_{\perp}$ Bundles on the Heterotic Side}

\subsubsection{Spectral Cover Construction}

To start, let us briefly review the spectral cover construction of $SU(5)_{\perp}$ bundles on the Heterotic side and the chiral spectrum that they induce \cite{Friedman:1997yq,Donagi:2004ia,Blumenhagen:2006wj}.  We start with an elliptically fibered Calabi-Yau 3-fold ${\cal{Z}}_H$ with section $\sigma_H$ over a two-dimensional base $S_2$.  We use $\pi_H$ to denote the projection map
\begin{equation}\pi_H:{\cal{Z}}_H\rightarrow S_2\,.\end{equation}
Because it has a section, $\sigma_H$, we can specify ${\cal{Z}}_H$ by a Weierstrass equation.  For our purposes, it will be convenient to realize the elliptic fiber as a curve inside $\mathbb{P}^2_{1,2,3}$ with coordinates $[v,x,y]$ and hence ${\cal{Z}}_H$ as a submanifold of a $\mathbb{P}^2_{1,2,3}$ fibration over $S_2$ that is given by the Weierstrass equation
\begin{equation}y^2 = x^3 + f_4 x v^4 + g_6v^6\,.\end{equation}
Here $v$, $x$, and $y$ are sections of ${\cal{O}}(\sigma_H)$, ${\cal{O}}(\sigma_H^2)\otimes K_{S_2}^{-2}$ and ${\cal{O}}(\sigma_H^3)\otimes K_{S_2}^{-3}$, respectively, meaning that $f_4$ and $g_6$ are sections of $K_{S_2}^{-4}$ and $K_{S_2}^{-6}$.  Divisor classes on ${\cal{Z}}_H$ include pullbacks $\pi_H^*\delta$ of curves $\delta$ inside $S_2$ as well as the zero section, $\sigma_H$, which satisfies the relation
\begin{equation}\sigma_H\cdot_{ {\cal{Z}}_H} (\sigma_H+\pi_H^*c_1) = 0\,.\end{equation}
Note that here we follow standard notation and use $c_1$ as shorthand for $c_1(S_2)$.
Following Friedman, Morgan, and Witten \cite{Friedman:1997yq}, one can construct an $SU(5)_{\perp}$ bundle on ${\cal{Z}}_H$ by specifying a spectral cover ${\cal{C}}_H\subset {\cal{Z}}_H$
\begin{equation}{\cal{C}}_H:\quad a_5 xy + a_4 x^2v + a_3 yv^2 + a_2 xv^3 + a_0v^5\,.
\label{speccoverdef}\end{equation}
Here, $a_0$ is a section of some bundle on $S_2$ that we specify by ${\cal{O}}(\eta)$.  The remaining $a_m$, then, are sections of ${\cal{O}}(\eta - mc_1)$.

As a divisor inside ${\cal{Z}}_H$, ${\cal{C}}_H$ is in the class
\begin{equation}{\cal{C}}_H = 5\sigma_H + \pi_H^*(6c_1-t)\,.\end{equation}
We will also make use of the projection
\begin{equation}p_H:{\cal{C}}_H\rightarrow S_2\,.\end{equation}

The $SU(5)_{\perp}$ bundle specified by ${\cal{C}}_H$ can be further twisted by introducing a $(1,1)$-form, $\gamma_H$, on ${\cal{C}}_H$.  The Chern classes of the bundles specified by ${\cal{C}}_H$ and $\gamma_H$ can be found in \cite{Friedman:1997yq,Donagi:2004ia,Blumenhagen:2006wj}.  For our immediate purposes, it is important to note that the first Chern class vanishes, meaning that we have a true $SU(5)_{\perp}$ bundle, precisely when the twist satisfies the "traceless" condition
\begin{equation}p_{ {\cal{C}}_H*}\gamma=0\,.\end{equation}
In general, there are not many choices for $\gamma$.  When the sections $a_m$ are reasonably generic, for instance, the only nontrivial divisors inside ${\cal{C}}_H$ arise from intersections with the section, $\sigma_H$, and pullbacks $\pi_H^*\delta$ of curves $\delta$ inside $S_2$.  Anything of the form $\pi_H^*\delta$ is pure trace so to construct a "traceless" $\gamma$, we must start with the section $\sigma_H$.  This can be done as
\begin{equation}\begin{split}\gamma_H &= (\sigma_H\cdot {\cal{C}}_H) - p_{ {\cal{C}}_H}^*p_{ {\cal{C}}_H*}(\sigma_H\cdot {\cal{C}}_H) \\
&= {\cal{C}}_H \cdot \left[5\sigma_H - \pi_H^*(c_1-t)\right]\end{split}\end{equation}

The $SU(5)_{\perp}$ bundle described above breaks one $E_8$ down to a subgroup that we call $SU(5)_{\rm GUT}$.  Under the breaking, the $E_8$ adjoint gives rise to charged fields in both the $\mathbf{10}$ and $\mathbf{\overline{5}}$ representations of $SU(5)_{\rm GUT}$ (and their conjugates) whose zero modes are counted by cohomology groups of the twisted $SU(5)_{\perp}$ bundle.  For our purposes, we are interested in the net chirality, which is counted by indices whose computation localizes on "matter curves" inside ${\cal{C}}_H$.  More specifically, the net chirality of a given representation is evaluated by identifying a suitable "matter curve" inside ${\cal{C}}_H$ and integrating the twist $\gamma_H$ over it \cite{Donagi:2004ia,Blumenhagen:2006wj}.
In the case of $\mathbf{10}$'s, for instance, the matter curve is just the intersection of ${\cal{C}}_H$ with the zero section
\begin{equation}\Sigma_{10,H} = \sigma_H\cdot_{ {\cal{Z}}_H } {\cal{C}}_H\,.\end{equation}
This leads to the standard chirality relation, which can be evaluated by a simple intersection computation
\begin{equation}\begin{split}\chi(\mathbf{10}) &= \int_{\Sigma_{10,H}}\gamma_H \\
&= \gamma_H\cdot_{ {\cal{C}}_H } \Sigma_{10,H} \\
&= {\cal{C}}_H \cdot_{ {\cal{Z}}_H } \left[5\sigma_H - \pi_H^*(c_1-t)\right]\cdot_{ {\cal{Z}}_H }\sigma_H \\
&= -(c_1-t)\cdot_{S_2} (6c_1-t)\,.
\end{split}\end{equation}

Counting the chirality of $\mathbf{\overline{5}}$'s, on the other hand, is slightly trickier.  To define the $\mathbf{\overline{5}}$ "matter curve", one starts by defining the involution $\tau$ that takes $y\rightarrow -y$ while leaving $x$ and $v$ invariant.  The curve $\Sigma_{\mathbf{\overline{5}},H}$ is now obtained as a particular component of the locus ${\cal{C}}_H\cap \tau {\cal{C}}_H$.  To describe which one, consider the full intersection ${\cal{C}}_H\cap \tau {\cal{C}}_H$, which is specified by the equations
\begin{equation}v\left[a_0 v^4 + a_2 v^2x + a_4\right]=0\qquad y\left[a_3 v^2 + a_5x\right]=0\,.\end{equation}
This includes several components
\begin{equation}\begin{split}0&=v = y\\
0&=v=a_3v^2+a_5x\\
0&=y=a_0v^4+a_2v^2x+a_4\\
0&=a_3v^2+a_5x=a_0v^4+a_2v^2x+a_4\,.
\end{split}\end{equation}
The last one is the curve $\Sigma_{\mathbf{\overline{5}},H}$ \cite{Donagi:2004ia,Blumenhagen:2006wj}.  Homologically, it is easy to see that it descends from a divisor inside ${\cal{Z}}_H$
\begin{equation}\begin{split}\Sigma_{\mathbf{\overline{5}}_H} &= {\cal{C}}_H\cdot {\cal{Z}}_H {\cal{C}}_H - [v]\cdot_{ {\cal{Z}}_H } [a_5 x] - [y]\cdot_{ {\cal{Z}}_H } [a_4] - [v]\cdot_{ {\cal{Z}}_H } [y] \\
&={\cal{C}}_H\cdot_{ {\cal{Z}}_H} \left[ {\cal{C}}_H - 3(\sigma+\pi_H^*c_1)-\sigma_H\right]\,.
\end{split}\end{equation}
The net chirality of $\mathbf{\overline{5}}$'s can now be computed as
\begin{equation}\begin{split}\chi(\mathbf{\overline{5}}) &= \gamma_H\cdot_{ {\cal{Z}}_H }\Sigma_{\mathbf{\overline{5}},H}\\
&= {\cal{C}}_H\cdot \left[5\sigma_H - \pi^*_H(c_1-t)\right]\cdot \left[{\cal{C}}_H - 3(\sigma_H+\pi^*_Hc_1)-\sigma_H\right] \\
&= -(6c_1-t)\cdot_{S_2}(c_1-t)\,.
\end{split}\end{equation}
That we get the same result as the net chirality of $\mathbf{10}$'s was guaranteed by the fact the spectral cover construction has produced a bundle $V$ with $c_1(V)=0$ \cite{Friedman:1997yq,Donagi:2004ia,Blumenhagen:2006wj}.  Ultimately, this property is connected to the "traceless" requirement on $\gamma_H$.

\subsubsection{del Pezzo Construction}\label{subsubsec:dp}

An alternative description of $SU(5)_{\perp}$ bundles on the Heterotic side can be given by a construction involving del Pezzo surfaces.  For this, let us briefly recall the method of Friedman, Morgan, and Witten for constructing $E_8$ bundles from $dP_8$ surfaces \cite{Friedman:1997yq}.  The anti-canonical class of $dP_8$, which we refer to as $x_8$, is an elliptic curve that we will eventually identify with the elliptic fiber of ${\cal{Z}}_H$.  The orthogonal complement to $x_8$ inside $H_2(dP_8,\mathbb{Z})$ is generated by a set of classes that the authors of \cite{Hayashi:2008ba} termed $R_8$
\begin{equation}R_8 = \{C\in H_2(dP_8,\mathbb{Z})\,|\, C\cdot x_8=0,\,\,\, C^2=-2\}\,.
\end{equation}
Because a fixed element $C\in R_8$ has vanishing intersection with $x_8$, the line bundle ${\cal{O}}(C)$ on $dP_8$ restricts to a flat bundle there.  The full set $R_8$ therefore defines a collection of flat bundles on $x_8$.  Now, suppose we fiber $dP_8$ over some base surface, $S_2$.  As we do this, elements of $R_8$ will in general mix under the action of some monodromy group, which in turn interchanges the bundles.  These monodromies must preserve the intersection form, however, which in the case of $R_8$ is proportional to the Cartan matrix of $E_8$.  The isomorphism between $R_8$ and the set of roots of the Lie algebra $E_8$ means that any monodromy must act via an element of the Weyl group.  What we have really obtained, then, is an $E_8$ bundle on the fiberwise restriction of the $dP_8$-fibration to $x_8$, and hence to an elliptic fibration over $S_2$.

To visualize this, it is often helpful to work not with the roots $R_8$ but instead the set of exceptional lines $I_8$ \cite{Hayashi:2008ba}
\begin{equation}I_8 = \{\ell\in H_2(dP_8,\mathbb{Z})\,|\,\ell\cdot x_8 = 1,\,\,\,\ell^2=-1\}\,.\end{equation}
Elements $\ell\in I_8$ are in 1-1 correspondence with elements $C\in R_8$ via{\footnote{The authors of \cite{Hayashi:2008ba} write this relation as $\ell=x_8+C$.  We choose $x_8-C$ because we will want to think of some $\ell$'s as proper transforms representatives of $x_8$ that meet a singular point where some $C$'s collapse.}}
\begin{equation}\ell = x_8 - C\,.\end{equation}
Any element $\ell\in I_8$ intersects $x_8$ in a single point, $p$, which defines the flat bundle ${\cal{O}}(p-p_0)${\footnote{Here, $p_0$ is the marked point on the elliptic curve $x_8$.}}.

For comparison to F-theory, it is useful to realize this picture inside $dP_9$ rather than $dP_8$ \cite{Curio:1998bva,Donagi:2008ca,Hayashi:2008ba}, which we think of as an elliptic fibration over a base $\mathbb{P}^1$.  We identify the elliptic fiber with the anti-canonical class, $x_9$, and the base with the exceptional curve $e_9$.  The sets $R_8$ and $I_8$ are realized in this setting as
\begin{equation}R_8 = \{C\in H_2(dP_9,\mathbb{Z})\,|\, C\cdot x_9= C\cdot e_9 = 0,\,\,\, C^2=-2\}\,.\end{equation}
and
\begin{equation}I_8 = \{\ell\in H_2(dP_9,\mathbb{Z})\,|\,\ell\cdot x_9 = 1,\,\,\, \ell\cdot e_9=0,\,\,\ell^2=-1\}\,.\end{equation}

For model-building, we are not interested in generic $E_8$ bundles but rather proper subbundles.  For this, the $dP_9$ fibration must be such that the monodromies mix $I_8$ only according to a proper subgroup of the Weyl group of $E_8$.  This requires singling out some of the lines from the others, which can be accomplished by requiring the $dP_9$ to exhibit a local K3 singularity.  If the $dP_9$ exhibits an $SU(5)_{\rm GUT}$ singularity over $S_2$, for instance, then the exceptional
lines corresponding to roots of $SU(5)_{\rm GUT}$ will not mix with those of the commutant $SU(5)_{\perp}$, effectively reducing the monodromy group.  The orbits of $I_8$ under the monodromy group action are in 1-1 correspondence with the $SU(5)_{\rm GUT}\times SU(5)_{\perp}$ representations that descend from the $E_8$ adjoint
 \begin{equation}\mathbf{248}\rightarrow (\mathbf{24},\mathbf{1})\oplus (\mathbf{1},\mathbf{24})\oplus \left[(\mathbf{10},\mathbf{5})\oplus cc\right]\oplus \left[(\mathbf{\overline{5}},\mathbf{10})\oplus cc\right]\,.\end{equation}
 To determine the $SU(5)_{\perp}\subset E_8$ subbundle that is specified by the fibration, it is enough to focus on one set of five lines that transform as a $\mathbf{5}$ under $SU(5)_{\perp}${\footnote{It is impossible to distinguish different  components of the $\mathbf{10}$ in $(\mathbf{10},\mathbf{5})$ without resolving the $SU(5)_{\rm GUT}$ singularity.}}.  Above any generic point in $S_2$, this union of lines will intersect $x_9$ at five points.  Fibering over $S_2$, this intersection is promoted to a 5-fold cover of $S_2$ which is nothing other than the Heterotic spectral cover, ${\cal{C}}_H$.  In this way, we recover the spectral cover construction of the previous subsection.

\subsubsection{Connecting the Spectral Cover and del Pezzo Constructions}\label{subsubsec:scdp}

To illustrate this, let us consider an explicit $dP_9$-fibration over a complex surface $S_2$.  Because the $dP_9$ is elliptically fibered, we can view this as an elliptic fibration over a 3-dimensional base $B_3$ which takes the form of a $\mathbb{P}^1$ fibration over $S_2$.  To specify $B_3$, it is enough to specify a line bundle ${\cal{T}}$ on $S_2$ with $c_1({\cal{T}})=t$ by which to twist the fibration.  In this case, we can think of $B_3=\mathbb{P}({\cal{O}}\oplus {\cal{T}})$.  Following \cite{Donagi:2008ca}, we will use $\rho$ to denote the $\mathbb{P}^1$ fibration
\begin{equation}\rho:B_3\rightarrow S_2\,.\end{equation}
Divisor classes in $B_3$ include pullbacks $\rho^*\delta$ of curves $\delta$ inside $S_2$ as well as a new class $r$ that satisfies
\begin{equation}r(r+\rho^*t)=0\,.\end{equation}
This relation can be understood because the projective coordinates, $Z_1$ and $Z_2$, on the $\mathbb{P}^1$ fiber above a fixed point on $S_2$ are promoted to sections of the bundles $r$ and $r+\rho^*t$ as we move along $S_2$.  One can compute the Chern classes of $B_3$ by adjunction, with the result
\begin{equation}c_1(B_3) = \rho^*(c_1+t)+2r\qquad c_2(B_3)=\rho^*c_2+\rho^*c_1\cdot(\rho^*t+2r)\qquad c_3(B_3)=\rho^*c_2\cdot(2r+\rho^*t)\,,\end{equation}
where $c_a:=c_a(S_2)$ for $a=1,2.$

We now proceed to construct a full $dP_9$-fibration, $Y_4$.  Following \cite{Donagi:2008ca}, we will use $\pi$ to denote the projection map of the elliptic fibration
\begin{equation}\pi:Y_4\rightarrow B_3\end{equation}
and $p$ to denote the projection map of the $dP_9$-fibration
\begin{equation}p:Y_4\rightarrow S_2\,.\end{equation}
To describe $Y_4$, we again realize the elliptic fiber as a submanifold of $\mathbb{P}^2_{1,2,3}$ with coordinates $[v,x,y]$ and write a Weierstrass equation \cite{Hayashi:2008ba}
 \begin{equation}\begin{split}y^2 &= x^3 + f_4Z_1^4 x v^4 + g_6 v^6 Z_1^6 \\
&\qquad + Z_2\left[b_0 (vZ_1)^5 + b_2(vZ_1)^3x + b_3(vZ_1)^2y + b_4(vZ_1)x^2 + b_5xy\right]\,.
\end{split}\end{equation}
Here, $v,x,y$ are sections of
${\cal{O}}(\sigma)$, ${\cal{O}}^2\bigl(\sigma+\pi^*r +p^*c_1\bigr)$, and ${\cal{O}}^3\bigl(\sigma+\pi^*r +p^*c_1\bigr)$, respectively, while $Z_1$ and $Z_2$ are sections of $\pi^*r$ and $\pi^*r+p^*t$.  With this choice, $f_4$ and $g_6$ are sections of $K_{S_2}^{-4}$ and $K_{S_2}^{-6}$, respectively, meaning that the submanifold $Z_2=0$ is precisely the Heterotic 3-fold ${\cal{Z}}_H$.

According to our previous discussion, the $dP_9$-fibration above specifies an $SU(5)_{\perp}$ bundle on ${\cal{Z}}_H$ whose structure we can deduce by studying the collection of five exceptional lines comprising a $\mathbf{5}$ of $SU(5)_{\perp}$.  As discussed in \cite{Hayashi:2008ba}, the union of these lines is simply the divisor
\begin{equation}{\cal{C}}:\quad b_0 (vZ_1)^5 + b_2(vZ_1)^3x + b_3(vZ_1)^2y + b_4(vZ_1)x^2 + b_5xy\,.\end{equation}
The intersection of ${\cal{C}}$ with ${\cal{Z}}_H$ is nothing other than the Heterotic spectral cover, provided we identify the restriction to ${\cal{Z}}_H$ of the sections $b_m$ here with the sections $a_m$ of \eqref{speccoverdef}.  In this case, the $b_m$ are sections of $p^*((6-m)c_1-t)$, whose restriction to ${\cal{Z}}_H$ is $\pi_H^*((6-m)c_1-t)$.  This leads to the standard identification
\begin{equation}\eta = 6c_1-t\,.\end{equation}

Note that this $dP_9$ construction has reproduced the spectral bundles of the previous subsection without the additional twist, $\gamma$.  Apparently, the geometry of $dP_9$ alone is insufficient to capture this additional structure.
As we now review, the $dP_9$ in this construction is "physical" in the sense that it corresponds to part of an F-theory compactification that is determined by bundle data of the Heterotic dual.  The twist of the Heterotic bundle, however, adds a new structure to F-theory that is not encoded in pure geometry; it adds $G$-flux \cite{Curio:1998bva}\cite{Donagi:2008ca,Hayashi:2008ba}.

\subsection{Heterotic/F-theory Duality Map}

As is well-established by now, the Heterotic compactification described above admits a dual description in F-theory as an elliptically-fibered Calabi-Yau 4-fold, $Z_4$, whose base $B_3$ is a $\mathbb{P}^1$-fibration over a 2-fold $S_2$.  We will use the same realization for $B_3$ as in section \ref{subsubsec:scdp} as well as the same notation for various projection maps.

The F-theory 4-fold, $Z_4$, is specified by a Weierstrass equation
\begin{equation}y^2 = x^3 + fxv^4 + gv^6\,,\end{equation}
where here, because we want a Calabi-Yau, $f$ and $g$ should be sections of $K_{B_3}^{-4}$ and $K_{B_3}^{-6}$.  To achieve this, we can take $f$ and $g$ to be homogeneous polynomials in the $Z_i$ of degrees 8 and 12, respectively
\begin{equation}f = \sum_{m=0}^8 f_m Z_1^m Z_2^{8-m}\qquad g = \sum_{n=0}^{12} g_n Z_1^n Z_2^{12-n}\,.\end{equation}
The relation to Heterotic arises in the stable degeneration limit \cite{Morrison:1996na,Morrison:1996pp}, in which the K3 fibers split into a pair of $dP_9$'s glued together along a common elliptic curve.  To describe this limit, it is helpful to write
\begin{equation}\begin{split}y^2 -x^3 =& xv^4\left[Z_2^4\sum_{m=0}^3 f_m Z_1^m Z_2^{4-m}\right] + v^6Z_2^6\sum_{n=0}^5 g_n Z_1^n Z_2^{6-n} \\
& + xv^4f_4Z_1^4Z_2^4 + v^6g_6 Z_1^6Z_2^6 \\
& + xv^4\left[Z_1^4\sum_{m=0}^3 f_{8-m} Z_1^{4-m}Z_2^n\right] + v^6Z_1^6\sum_{n=0}^5 g_{12-n}Z_1^{6-n}Z_2^n\,.
\end{split}\end{equation}
Near $Z_1=0$, we can set $Z_2=1$ and the geometry is described by one $dP_9$ fibration
\begin{equation}y^2 = x^3 + xv^4 \sum_{m=0}^4 f_m z_1^m + v^6\sum_{n=0}^6 g_n z_1^n\,,\end{equation}
while we have a similar story near $Z_2=0$.  The two $dP_9$'s are glued along an elliptically-fibered 3-fold given by
\begin{equation}y^2 = x^3 + x v^4 f_4 + v^6 g_6\,.\end{equation}
Note that $f_4$ and $g_6$ are sections of $K_{S_2}^{-4}$ and $K_{S_2}^{-6}$, respectively, so this is a Calabi-Yau that is identified with the Heterotic 3-fold ${\cal{Z}}_H$.

That we have effectively two $dP_9$'s in this limit reflects the fact that the Heterotic compactification is specified not just by ${\cal{Z}}_H$ but by a pair of $E_8$ bundles.  The geometry of each $dP_9$ on the F-theory side, as specified by the suitable sections $f_m$ and $g_n$, is determined by the spectral data of a Heterotic bundle according to the del Pezzo construction of section \ref{subsubsec:dp}.

We are interested in only one spectral bundle so we will focus on one of the two $dP_9$'s.  This will suffice for any considerations, such as the spectrum of chiral matter fields from one $E_8$, which depend only on one of the bundles on the Heterotic side and, correspondingly, involve only local data near $Z_1=0$ on the F-theory side.  Note that issues such as anomaly cancellation on the Heterotic side, which depend on both bundles, map to global questions on the F-theory side that will require us to move beyond the geometry of a single $dP_9$.

\subsubsection{$\gamma_H$ and $G$-fluxes}

On the Heterotic side, chirality has its origin in the twist $\gamma_H$ of the $SU(5)_{\perp}$ spectral bundle \cite{Donagi:2004ia,Blumenhagen:2006wj}.  In F-theory, we expect chirality to originate from $G$-flux so one expects a mapping between the two.  In the context of Heterotic/F-theory duality, this mapping has been described in many places in the literature before \cite{Curio:1998bva}\cite{Donagi:2008ca,Hayashi:2008ba}.  Here, we review the relevant facts.

The connection between del Pezzo surfaces and the Heterotic spectral cover is often described by the cylinder map \cite{Curio:1998bva}.  The "cylinder", denoted by $R$, is a union of lines in $dP_8$ which can also be thought of as a union of lines in $dP_9$ that do not intersect the exceptional curve $e_9$.  As we saw in section \ref{subsubsec:scdp}, the spectral cover ${\cal{C}}_H$ arises as the intersection of a distinguished subset of the "cylinder" with ${\cal{Z}}_H$.  For this reason, it is natural to consider a projection map
\begin{equation}p_R:R\rightarrow {\cal{C}}_H\,.\end{equation}
When we add the $(1,1)$-form $\gamma_H$ on ${\cal{C}}_H$, then, it is natural to suppose that the $G$-flux is obtained by pulling it back to $R$.  More specifically, one has \cite{Curio:1998bva}\cite{Donagi:2008ca,Hayashi:2008ba}
\begin{equation}G = i_* p_R^*\gamma\label{oldG}\,,\end{equation}
where $i$ is the inclusion map of $R$ into the 4-fold $Y_4$
\begin{equation}i:R\rightarrow Y_4\,.\end{equation}


\subsubsection{Supersymmetry and Lorentz Invariance}
\label{subsubsec:susyli}

Two important conditions on $G$ are that it is supersymmetric and corresponds to a Lorentz invariant flux in F-theory.  Fluxes constructed via the cylinder map prescription above are modified to satisfy these conditions by adding a constant piece \cite{Hayashi:2008ba,Donagi:2009ra}
\begin{equation}G = i_*p_R^*\gamma_H - q G_0\,,
\label{Golddef}\end{equation}
where $G_0$ is the Poincare dual of $S_2$ inside $Y_4$ and $q$ is given by
\begin{equation}q = \gamma_H\cdot_{ {\cal{C}}_H} \sigma_H|_{ {\cal{C}}_H }\,.\end{equation}

Let us take a moment to recall where this comes from.  The condition of Lorentz invariance in F-theory is the statement that $G$ integrates to zero over any divisor inside $B_3$
\begin{equation}\int_D\,\pi_*G=0\qquad \text{ for any }D\in H_4(B_3,\mathbb{Z})\,.\end{equation}
We will typically identify $G$ with a surface in $Z_4$, in which case we can phrase this by saying that $G$ has vanishing intersection with any surface in $Z_4$ that coincides with a divisor in $B_3$.  Equivalently, this means that $G$ has vanishing intersection with $B_3$ itself.

Supersymmetry, on the other hand, is the statement that
\begin{equation}J\wedge G=0\,,\end{equation}
for any suitable K\"ahler form $J$.  Recall that K\"ahler forms on $Z_4$ are just pullbacks of K\"ahler forms on $B_3$.  What we mean by the above statement is that $G$, viewed as a surface, has vanishing intersection with any divisor on $Z_4$ that takes the form of an elliptic fibration over a 4-cycle inside $B_3$.  Of course, we should only consider K\"ahler forms inside the K\"ahler cone but the object \eqref{Golddef} was actually constructed so that the stronger condition holds \cite{Hayashi:2008ba}.

What we have essentially claimed is that $G$ has vanishing intersection with $B_3$ as well as any divisor that takes the form of an elliptic fibration over a 4-cycle in $B_3$.  This exhausts the set of divisors inside $Z_4$, so what we are saying is that $G$ is trivial in the homology of $Z_4$.  In fact, it is easy to see that this is basically the case.  The traceless condition on $\gamma_H$ inside ${\cal{Z}}_H$ means that $\iota_{H*}\gamma_H = nF$ where $F$ is the elliptic fiber class in ${\cal{Z}}_H$ and $\iota_H$ is the projection
\begin{equation}\iota_H:{\cal{C}}_H\rightarrow {\cal{Z}}_H\,.\end{equation}
The elliptic fiber class meets the zero section once so the constant of proportionality, $n$, is just $n=\gamma_H\cdot_{ {\cal{C}}_H }\sigma=q$.  Pulling back $\gamma_H$ by $p_R^*$ gives $q$ times the $dP_9$ fiber class of $Y_4$ or, in other words, $q$ times the Poincare dual of $S_2$.  Subtracting $qG_0$ from $\iota_*p_R^*\gamma_H$ therefore yields something that is trivial in the "naive" homology of $Z_4$.

Of course, $G$ is not really trivial because the "naive" homology of $Z_4$ does not take into account the effects of the resolution.  The cylinder will be sensitive to the resolution so the homology class of the first term in \eqref{Golddef} will be modified relative to the second.  The difference between the two will involve some combination of resolved cycles.  The K\"ahler form is unmodified by the resolution because the new cycles should have zero volume so the supersymmetry condition remains intact.  Further, the resolved cycles have a leg on the torus so they are Lorentz invariant.

In the proposal of the next section, we will arrive at a slightly different prescription for the $G$-fluxes that does not involve precisely $p_R^*$.  Therefore, the "naive" homology class of the $G$-flux specfied by $\gamma_H$ will not be of the form of a $dP_9$ fiber class.  Nevertheless, one can use the same strategy to construct supersymmetric and Lorentz invariant fluxes; one can identify the naive class of $G$ inside the singular $Z_4$ and add a new contribution that is not sensitive to the resolution and such that the net $G$ is trivial in the "naive" homology of $Z_4$.  Such an additional constant term will always be implied in the $G$-fluxes that we construct.





\section{Spectral Divisor Construction}

We now turn to our general proposal for describing $G$-fluxes.  We will define an object ${\cal{C}}$ that we refer to as the "spectral divisor" of the F-theory 4-fold, $Z_4$, and define the $G$-flux by specifying a $(1,1)$-form $\gamma$ on ${\cal{C}}$.  With this definition, we will study the chiral spectrum \cite{Hayashi:2008ba},
\begin{equation}\chi(R)=\int_{ \hat{\Sigma}_R}G\,,\end{equation}
where $\hat{\Sigma}_R$ is the "matter surface" obtained by taking a matter curve, $\Sigma_R$, and combining it with the 2-cycle that degenerates there.  On general grounds, we will demonstrate that $\chi(R)$, which naively depends on details of the resolution, can be computed as
\begin{equation}\chi(R)=\int_{\tilde{\Sigma}_R\cdot_{ {\cal{C}} } p_{ {\cal{C}}}^*S_2}\gamma\,,
\label{chirality}\end{equation}
where $\tilde{\Sigma}_R$ is a suitably defined "dual matter surface" inside ${\cal{C}}$.

To motivate the construction, we start in the context of a $dP_9$ fibration, $Y_4$, over $S_2${\footnote{Working in $Y_4$, rather than an elliptically fibered Calabi-Yau 4-fold $Z_4$, makes comparison with Heterotic results easier because the Heterotic 3-fold, ${\cal{Z}}_H$, represents a nontrivial divisor class.}}.  There, our goal is to make the procedure \eqref{oldG} more concrete{\footnote{By this, we mean to give an explicit definition of the $G$-flux on the F-theory side.  The ambiguity we have in mind is related to the fact that there are potential contributions to $\gamma$ that have trivial restriction to the Heterotic spectral cover, ${\cal{C}}_H$, but nevertheless must be included for validity of the counting prescription that is intrinsic to F-theory.  One may view this as identifying the right object to use in place of $p_R^*\gamma_H$ for the purpose of practical computations in terms of naive intersection theory on the singular $Y_4$.}} and formulate it in such a way that it can be generalized to 4-folds that do not admit a global Heterotic dual.  We then use local intersection data near the singular locus to derive the chirality formula \eqref{chirality} in this setting.  Because we are careful not to use the $dP_9$-fibration or any global intersection data, the derivation naturally carries over to more general F-theory 4-folds.   After showing that \eqref{chirality} is equivalent to the standard Heterotic formulae when a Heterotic dual exists, we demonstrate a similar agreement in a more general setting.  After that, we comment on quantization rules for $G$-fluxes.

\subsection{Spectral Divisor when a Heterotic Dual exists}

We start with a $dP_9$-fibration $Y_4$ given by
\begin{equation}y^2 = x^3 + x v^4 f_4 Z_1^4 + g_6 v^6 Z_1^6 + Z_2\left[b_0 (vZ_1)^5 + b_2(vZ_1)^3x + b_3 (vZ_1)^2y + b_4(vZ_1)x^2 + b_5 xy\right]\,,\label{Y4def}\end{equation}
and define a "spectral divisor" ${\cal{C}}$ inside $Y_4$ by
\begin{equation}{\cal{C}}:\quad b_0 (vZ_1)^5 + b_2(vZ_1)^3x + b_3 (vZ_1)^2y + b_4(vZ_1)x^2 + b_5 xy
\label{specdivdef}\end{equation}
along with a projection
\begin{equation}p_{ {\cal{C}}}:{\cal{C}}\rightarrow B_3\,.\end{equation}
This is nothing other than the union of lines in section \ref{subsubsec:dp} fibrered over $S_2$.  In addition to this interpretation, however, we want to think of this roughly as specifying an $SU(5)$ bundle on $Y_4$, interpreting ${\cal{C}}$ as a spectral cover along the lines of \cite{Friedman:1997yq}{\footnote{Of course, this is a very special bundle because $Y_4$ is singular and the divisor ${\cal{C}}$ has nontrivial intersection with the singular locus}}.  This bundle can be twisted by specifying a $(1,1)$-form $\gamma$ on ${\cal{C}}$.  Just as the restriction of ${\cal{C}}$ to the Heterotic 3-fold ${\cal{Z}}_H$ at $Z_2=0$ is ${\cal{C}}_H$, the restriction of $\gamma$ will yield a $(1,1)$-form on ${\cal{C}}_H$ that is naturally identified with $\gamma_H$.  If we think of ${\cal{C}}$ as a stand-in for the cylinder, $R$, then $\gamma$ is a useful stand-in for $p_R^*\gamma_H$.  Defining $i_C$ as the embedding
\begin{equation}i_C:{\cal{C}}\rightarrow Y_4\,,\end{equation}
we therefore propose that the $G$-flux is specified by the twist, $\gamma$, according to
\begin{equation}G = i_{C*}\gamma\,.\end{equation}
As we shall see, this seems different than saying that $G$ comes from $i_*p_R^*\gamma$.  Viewed as a curve inside ${\cal{Z}}_H$, $\gamma$ (or more properly $i_{H*}\gamma$) is a multiple of the elliptic fiber class.  This suggests that $i_*p_R^*\gamma$ should be a multiple of the $dP_9$ fiber class, which is the Poincare dual to $S_2$.  As reviewed in section \ref{subsubsec:susyli}, this plays a role in determining the term that must be added to make the total flux supersymmetric and Lorentz invariant.  Once we impose the proper analog of the "traceless" condition on $\gamma$, we will see that an additional nonzero contribution with trivial restriction to ${\cal{Z}}_H$ must be added.  This contribution will not play any role in the formulae for counting chiral matter on the Heterotic side but will be crucial for getting the proper results on the F-theory side{\footnote{Because $i_*p_R^*\gamma$ is naively traceless as well, one may view our construction with traceless $\gamma$ as the "right" way to homologically describe this flux in F-theory in a way that is amenable to practical computations.}}.

The naive homological class of ${\cal{C}}$ inside $Y_4$ is simply
\begin{equation}{\cal{C}} = 5(\sigma + p^*c_1 + \pi^*r) + p^*(c_1-t)\,.\end{equation}
This is naive because ${\cal{C}}$ interesects the singular locus, which is found at $z=x=y=0$ and sits in the class
\begin{equation}{\cal{S}}_{\text{sing}} = (\sigma + p^*c_1 + \pi^*r)\cdot \pi^*r\,.\end{equation}

Because ${\cal{C}}$ intersects the singular locus, its actual homological class is somewhat ambiguous and depends on the resolution, after which it should be replaced by its proper transform.  In general, we will use ${\cal{C}}$ as a tool for constructing fluxes and counting chiral matter so these details will be important.  We will see, however, that it will be possible to reformulate the computation in such a way that it becomes unnecessary to worry about such things.  Only some well-known features of intersections in the local geometry near singular loci, which follow from the identification of ${\cal{C}}$ with the fibration of exceptional lines in $dP_9$ over $S_2$, will be needed.

\subsection{The General Chirality Formula}
\label{subsec:chirform}

We now turn to a discussion of the chirality formula.  We will work in the $dP_9$ setting for concreteness and to allow us to use the language of exceptional lines.  We are careful to make use only of local intersection data near the singular locus, though, which will allow the result to generalize.

In the case of a $dP_9$-fibration, ${\cal{C}}$ describes a sum of 5 lines in the $dP_9$ fibered over $S_2$.  To be definite, let us fix a point on $S_2$ and look at the restriction of ${\cal{C}}$ to the $dP_9$ fiber there.  Each line is in a class of the form $\ell^{(i)} = x_8 - C^{(i)}$ where the $C^{(i)}$ with $i=1,\ldots,5$ are five curves inside $dP_9$ with self-intersection -2 that transform as a $\mathbf{5}$ of the $SU(5)_{\perp}$ part of $E_8$.  The intersection matrix of the $C^{(i)}$ is well-known
\begin{equation}C^{(i)}\cdot C^{(j)} = -1 - \delta^{ij}\qquad C^{(i)}\cdot x_8=0\,.\end{equation}
Intersections of the lines $\ell^{(i)}$ with the $C^{(j)}$ is therefore also known
\begin{equation}\ell^{(i)}\cdot C^{(j)} = 1 + \delta^{ij}\,.\end{equation}
We have $\sum_{i=1}^5 C^{(i)}=0$ so that $\sum_{i=1}^5 \ell^{(i)} = 5x_8 = 5(x_9+e_9)$, consistent with the fact that, modulo ramification curves, ${\cal{C}}\sim 5(\sigma + \pi^*r)$.

In general, the singularity type enhances from $SU(5)$ to something larger when a suitable combination of $C^{(i)}$'s degenerate over a curve or point in $S_2$.  Let us denote a combination $C_{R,\,deg}$ that degenerates along a matter curve $\Sigma_R$ by
\begin{equation}C_{R,\,deg} = \beta_{R,a} C^{(a)}\,.\end{equation}
The class $\beta_{R,a}C^{(a)}$ inside $dP_9$ is represented by an effective curve that is "localized" at the singular locus.  This means that the intersection of $\beta_{R,a} C^{(a)}$ with the $\ell^{(i)}$'s that reside in ${\cal{C}}$, while computed above using the geometry of $dP_9$, is completely determined by the local geometry near the region of singular enhancement so does not depend on the global $dP_9$ structure.  This will be important in what follows.

We now turn to the chirality formula for fields living on the matter curve $\Sigma_R$
\begin{equation}\chi(R) = \int_{\hat{\Sigma}_R}G\,.\end{equation}
Since we have specified $G$ with a $(1,1)$-form inside ${\cal{C}}$, we rewrite this as
\begin{equation}\chi(R) = \int_{ {\cal{C}}\cdot \hat{\Sigma}_R}\gamma\,.
\label{chirform}\end{equation}
To describe the curve ${\cal{C}}\cdot\hat{\Sigma}_R$ over which we must integrate $\gamma$, it is helpful to explicitly display the fibration structure of ${\cal{C}}$ and $\hat{\Sigma}_R$
\begin{equation}{\cal{C}}\cdot \hat{\Sigma}_R = \left[\begin{pmatrix}\ell^{(i)} \\ \downarrow \\ S_2\end{pmatrix}\right]\cdot \left[\sum_a  \begin{pmatrix}\beta_{R,a}C^{(a)} \\ \downarrow \\ \Sigma_R\end{pmatrix}\right]\,.\label{Cmattsurf}\end{equation}
Evaluating this intersection is not completely straightforward.  For starters, the notation $(\ell^{(i)}\rightarrow S_2)$ is a bit schematic and neglects the fact that the $\ell^{(i)}$ are typically connected by monodromies and hence not individually well-defined.  This is problematic because we cannot really treat the $\ell^{(i)}$ all on an equal footing in \eqref{Cmattsurf}.  For a given choice of $a$, $\ell^{(i)}\cdot C^{(a)}$ depends on whether or not $i=a$.

To make headway, we first note that, because $\ell^{(i)}$ and $C^{(a)}$ intersect transversally, we can perform the intersection $\ell^{(i)}\cdot \beta_{R,a}C^{(a)}$ upstairs and fiber the result over $\Sigma_R$ to obtain
\begin{equation}{\cal{C}}\cdot \hat{\Sigma}_R = \sum_{i,a}\begin{pmatrix}\ell^{(i)}\cdot \beta_{R,a}C^{(a)} \\ \downarrow \\ \Sigma_R\end{pmatrix}\end{equation}

The specific combination $\beta_{R,a}C^{(a)}$ is an effective curve in the resolution that is "localized" on the matter curve $\Sigma_R$ so, in the limit that the resolution is turned off, the intersection happens in $p_{\cal{C}}^*\Sigma_R${\footnote{This is justified because what we mean by ${\cal{C}}$ is actually its proper transform under the resolution.  The curve ${\cal{C}}\cdot\hat{\Sigma}_R$ over which we integrate $\gamma$, then, meets the resolved cycle only at isolated points so we can reliably describe it in the limit that the resolution is turned off as long as we remember that it is a proper transform.}}.  The lift $p_{\cal{C}}^*\Sigma_R$ is a five-sheeted cover of $\Sigma_R$ inside ${\cal{C}}$ so the intersection numbers $\ell^{(i)}\cdot C^{(a)}$ assign individual multiplicities to each sheet as it appears in ${\cal{C}}\cdot \hat{\Sigma}_R$.  To determine these multiplicites, we turn to the local intersection data
\begin{equation}\ell^{(i)}\cdot C^{(a)} = 1 + \delta^{ia}\,.\label{localintdata}\end{equation}
The contribution from the "1" simply yields $(\sum_a\beta_{R,a}) p_{ {\cal{C}}}^*\Sigma$.  The $\delta^{ia}$ term is a bit trickier because it only gets contributions from sheets with $i=a$.  To describe this, we define the object $\tilde{\Sigma}_R$
\begin{equation}\tilde{\Sigma}_R = \sum_a \beta_{R,a} \begin{pmatrix}\ell^{(a)} \\ \downarrow \\ \Sigma_R\end{pmatrix}\,.
\end{equation}
Even though this has nothing to do with the matter surface \emph{per se}, we will abuse language and call this the "dual matter surface".  The definition above is rather schematic, as it requires us to isolate some combination of the $\ell^{(a)}$'s from the others.  Due to monodromies, this is generically not possible in ${\cal{C}}$.  When we restrict ${\cal{C}}$ to $p_{\cal{C}}^*\Sigma_R$, though, the degeneration of the combination $\beta_{R,a}C^{(a)}$ will cause the specific combination $\beta_{R,a}\ell^{(a)}$ to split off from the rest of ${\cal{C}}$.  We will see in detail how this works in examples below.

Returning to the $\delta^{ia}$ contribution to \eqref{localintdata}, we see that it corresponds to one copy of $\Sigma_R$ inside $p_{ {\cal{C}}}^*\Sigma_R$ for each sheet of the combination $\tilde{\Sigma}_R$.  Precisely the same thing is computed by the intersection $\tilde{\Sigma}_R\cdot_{ {\cal{C}} }p_{ {\cal{C}}}^*S_2$, ie by intersecting $\tilde{\Sigma}_R$ in $Y_4$ with an elliptic fibration over $S_2$.  This is because $x_9\cdot\ell^{(a)}=1$ for all $a$ and leads us to the result
\begin{equation}\chi(R) =\left(\sum_a \beta_{R,a}\right) \int_{ p_{\cal{C}}^*\Sigma_R}\gamma + \int_{ \tilde{\Sigma}_R\cdot_{ {\cal{C}} }p_{\cal{C}}^*S_2}\gamma\,.\end{equation}
In principle, this gives us a means of computing the spectrum for arbitrary choices of $\gamma$.  We would like to impose a further condition on $\gamma$, however, namely that it is "traceless"
\begin{equation}p_{{\cal C}*}\gamma=0\,.
\label{traceless}\end{equation}
This is motivated for several reasons.  First, it is necessary to ensure that $\gamma$ specifies a "traceless" flux $\gamma_H$ under the Heterotic/F-theory duality map.  Second, if we are naive about the singular locus, it will ensure that the twisted bundle on the Calabi-Yau 4-fold $Z_4$ that is specified by the spectral divisor construction has vanishing first Chern class, consistent with it being an $SU(5)_{\perp}$ bundle.  The most important reason to impose this condition, however, is that it guarantees a low energy spectrum without $SU(5)$ gauge anomalies.  This will follow because "tracelessness" allows us to connect to the Heterotic computation in the presence of a "traceless" $\gamma_H$, which is known to induce an anomaly-free spectrum \cite{Donagi:2004ia}.

Our final results, then, are to build a flux with a traceless $\gamma$ \eqref{traceless} and compute the net chirality via
\begin{equation}\chi(R) = \int_{ \tilde{\Sigma}_R\cdot_{ {\cal{C}} }p_{\cal{C}}^*S_2}\gamma\,.
\label{Ftheoryform}\end{equation}
Note that we did not have to make use of either the $dP_9$ fibration $p$ or any global intersection data to obtain this result.  Only the elliptic fibration and local intersection properties involving the matter surface were necessary.  This is crucial to justify generalizations to Calabi-Yau 4-folds without Heterotic duals.

\subsubsection{Connecting with the Heterotic Computation}

Before proceeding to examples, let us comment on how this approach is able to reproduce the standard Heterotic computation.  For this, it will be easiest to think in terms of the $dP_9$-fibration over $\Sigma_R$, $p^*\Sigma_R$, which we have avoided referring to thus far in order to keep the discussion as general as possible.  We will always assume a traceless $\gamma$ so that only the $\delta^{ia}$ piece of \eqref{localintdata} contributes in the end.  If this is the case, then we can actually replace ${\cal{C}}\cdot \hat{\Sigma}_R$ with
\begin{equation}\rho_R =\sum_{i,a}\begin{pmatrix}\beta_{R,i}\ell^{(i)}\cdot C^{(a)} \\ \downarrow \\ \Sigma_R\end{pmatrix}\,,\end{equation}
which will be easier to manipulate.  Because $\rho_R$ is contained in $p^*\Sigma_R$, we can compute \eqref{chirform} by restricting $\gamma$ to $p^*\Sigma_R$ and performing the integration there.  Because $\rho_R$ corresponds to isolated points in the $dP_9$ fiber sitting over $\Sigma_R$, any nonzero contribution to \eqref{chirform} will come via intersections with components of $\gamma$ that take the form of exceptional lines sitting over points in $\Sigma_R$.  This is important because it means that we do not care precisely where the intersection points $\ell^{(i)}\cdot C^{(a)}$ are located inside a given $\ell^{(i)}$.  Rather, we just care about how many intersection points we have in any given $\ell^{(i)}$.


The $\delta^{ia}$ piece of the sum $\sum_a \ell^{(i)}\cdot C^{(a)}$, which is the only one that contributes due to \eqref{traceless}, gives us 1 point on each $\ell^{(i)}$.
We can get the same result, though, by replacing the $\sum_a C^{(a)}$ with $x_9$ since $x_9\cdot \ell^{(i)}=1$.  This suggests that for practical purposes we should simply replace
\begin{equation}\sum_{i,a}\begin{pmatrix}\beta_i\ell^{(i)}\cdot C^{(a)} \\ \downarrow \\ \Sigma_R\end{pmatrix}\rightarrow \sum_i\begin{pmatrix}\beta_i\ell^{(i)} \\ \downarrow \\ \Sigma_R\end{pmatrix}\cdot \begin{pmatrix}x_9 \\ \downarrow \\ \Sigma_R\end{pmatrix}=\tilde{\Sigma}_R\cdot \begin{pmatrix}x_9 \\ \downarrow \\ \Sigma_R\end{pmatrix}\end{equation}
The important point now is that, in the $dP_9$ fibered case, there are two copies of $\Sigma_R$ corresponding to the two different sections of the $\mathbb{P}^1$-fibration, $B_3$.  One of these sits inside the copy of $S_2$ at $Z_1=0$ that we normally think of as the GUT divisor in the F-theory setting.  The other resides at $Z_2=0$, which is the Heterotic 3-fold ${\cal{Z}}_H$.  From the $dP_9$ perspective, these two choices are essentially equivalent, meaning that an elliptic fibration over either copy of $\Sigma_R$ can serve as a stand-in for $(x_9\rightarrow \Sigma_R)$.  Using either, the computation{\footnote{In order to phrase the result as a computation inside ${\cal{C}}$, it is necessary to promote the divisor $(x_9\rightarrow\Sigma_R)$ in $p^*\Sigma_R$ to the divisor $(x_9\rightarrow S_2)$ in $Y_4$.}}
\begin{equation}\chi(R) = \int_{ \tilde{\Sigma}_R\cdot_{ {\cal{C}}} D}\gamma \,, \qquad D = \begin{pmatrix}x_9 \\ \downarrow \\ S_2\end{pmatrix}\cdot {\cal{C}}\,.
\label{chiFhet}\end{equation}
yields an equivalent result.  Performing the computation at $Z_1=0$ corresponds to the F-theory formula \eqref{Ftheoryform}.  Performing the computation at $Z_2=0$, we will see that $\tilde{\Sigma}_R$ restricts to the usual Heterotic matter curve $\Sigma_{R,H}$ inside ${\cal{C}}_H$, $\gamma$ restricts to $\gamma_H$, and \eqref{chiFhet} reduces to the standard Heterotic result
\begin{equation}\chi(R) = \int_{ \Sigma_{R,H}}\gamma_H\,.\end{equation}
We will see this more explicitly in the examples that follow.

\subsection{Chiral spectrum from Inherited $G$-flux I -- Heterotic Case}

Let us now consider several examples.  We begin in this section by evaluating \eqref{Ftheoryform} in geometries with a Heterotic dual.  This will allows us to verify explicitly that \eqref{Ftheoryform} is consistent with the Heterotic formulae.  We will turn to more general 4-folds in the next subsection.

We focus here on a $dP_9$ fibration $Y_4$.  For generic ${\cal{C}}$, the $(1,1)$-forms available for constructing $\gamma$ are precisely those that descend from divisors inside $Y_4$.  Of those, anything of the form ${\cal{C}}\cdot \pi^*D$ is pure trace for $D$ a divisor in $B_3$.  An "inherited" $G$-flux that is traceless, then, takes the form
\begin{equation}\gamma = 5(\sigma\cdot {\cal{C}}) - p_C^*p_{C*} (\sigma\cdot {\cal{C}})\,.\end{equation}
A $G$-flux $G=\iota_*\gamma$ built from this particular $\gamma$ can be written as
\begin{equation}G = {\cal{C}}\cdot {\cal{G}}\,,\end{equation}
where
\begin{equation}{\cal{G}} = 5\sigma - p^*(c_1-t)\,.\end{equation}
Because $G$ descends from a divisor ${\cal{G}}$ inside $Y_4$, we can perform all chirality computations as intersections in $Y_4$ rather than in ${\cal{C}}$.  This means that once we identify the homology class of a "dual matter surface" $\hat{\Sigma}_R$, the chiral spectrum on the corresponding matter curve, $\Sigma_R$, can be determined as
\begin{equation}\chi(R) = {\cal{G}}\cdot \hat{\Sigma}_R\cdot \pi^*r\,.\label{Fcounting}\end{equation}

Finally, let us note that the naive cohomology class of $G$ inside $Y_4$ is given by
\begin{equation}G =  -p^*(c_1-t)\cdot \left[ p^*(6c_1-t) + 5 \pi^*r\right]\,,
\label{Gfluxex}\end{equation}
which, due to the $5\pi^*r$ term, is not quite the form of a $dP_9$-fibration.  This means that the "naive" class of $G$ inside $Y_4$ is not a $dP_9$ fiber class and hence the constant $G_0$ that must be added to ensure that the net $G$-flux is both supersymmetric and Lorentz invariant is not proportional to the Poincare dual of $S_2$.

\subsubsection{Counting $\mathbf{10}$'s}

Let's see how \eqref{Ftheoryform} looks when counting $\mathbf{10}$'s.  Restricting ${\cal{C}}$ to $p^*\Sigma_{10}$ can be done by plugging $b_5=0$ into the defining equation \eqref{specdivdef}.  When we do this, ${\cal{C}}$ factors into a quartic factor times a factor of $(Z_1v)$, which is in the class $\sigma + \pi^*r$.  What has happened here is that one of the $\ell^{(i)}$'s has degenerated into a reducible sum of the form $\ell^{(i)} = (\ell^{(i)}-e_9) + e_9$ as a result of the corresponding $C^{(i)}$ degenerating.  For this reason, the "dual matter surface" $\tilde{\Sigma}_{10}$ is simply $(Z_1v)=0$, $b_5=0$, which is in the class
\begin{equation}\tilde{\Sigma}_{10} = (\sigma + \pi^*r)\cdot \pi^*(c_1-t)\,.\end{equation}
Using the expression \eqref{Gfluxex} for the $G$-flux, we then find
\begin{equation}\chi(10) = \tilde{\Sigma}_{10}\cdot {\cal{G}}\cdot \pi^*r = -(6c_1-t)\cdot_{S_2}(c_1-t)\,,
\end{equation}
in agreement with the standard Heterotic result.

\subsubsection{Counting $\mathbf{\overline{5}}$'s}

Our $\mathbf{\overline{5}}$ "dual matter surface" corresponds to a pair $\ell^{(i)}$ and $\ell^{(j)}$ that satisfy $\ell^{(i)}+\ell^{(j)} = 0$ modulo $x_8$ above the $\mathbf{\overline{5}}$ matter curve.  As usual, we make use of the involution $\tau$ that sends $y\rightarrow -y$ because this also sends $\ell^{(i)}\rightarrow x_8 - \ell^{(i)}$.  The $\mathbf{\overline{5}}$ "dual matter surface" is then contained inside ${\cal{C}}\cap \tau {\cal{C}}$.  The full intersection is given by the equations
\begin{equation}(Z_1v)\left[b_0(Z_1v)^4 + b_2(Z_1v)^2x + b_4x^2\right]=0\qquad y\left[b_3(Z_1v)^2 + b_5 x\right]=0\,.
\label{CtauC}\end{equation}
The first thing we want to do is isolate the intersection of the two terms in brackets.  This class is given by
\begin{equation}{\cal{C}}\cdot {\cal{C}} - [y]\cdot[b_4x^2] - [Z_1v]\cdot [b_5x] - [y]\cdot [Z_1v]\,.
\end{equation}
Unlike the analogous Heterotic computation, however, what we have is reducible and a further component must be removed.  This component is the locus $x=Z_1=0$, which appears with multiplicity 4.  Subtracting that, we end up with
\begin{equation}\begin{split}\tilde{\Sigma}_{\mathbf{\overline{5}}} &= {\cal{C}}\cdot {\cal{C}} - [y]\cdot [b_4x^2] - [Z_1v]\cdot [b_5x] - [y]\cdot [Z_1v] - 4[x]\cdot[Z_1]\\
&\rightarrow 2(\sigma + \pi^*r)\cdot p^*(8c_1-3t) + p^*(3c_1-t)\cdot p^*(6c_1-t)\,.
\end{split}
\label{fivemattsurf}\end{equation}
It is easy to see explicitly from \eqref{CtauC} that what remains is a double cover of the $\mathbf{\overline{5}}$ matter curve inside $S_2$, $P_{\mathbf{\overline{5}}}=a_0a_5^2-a_2a_3a_5-a_4a_3^2$.  Using \eqref{fivemattsurf} to compute the net chirality of $\mathbf{\overline{5}}$'s, we again find agreement with the standard Heterotic formulae
\begin{equation}\chi(\mathbf{\overline{5}}) = \tilde{\Sigma}_{\mathbf{\overline{5}}}\cdot {\cal{G}}\cdot \pi^*r = -(6c_1-t)\cdot_{S_2}(c_1-t)\,.
\end{equation}

\subsubsection{More Detailed Comparison to Heterotic}

We can make the comparison with Heterotic more transparent by noting that it is possible to realize all of the objects that appear in the exact Heterotic computation as nontrivial homology classes inside $Y_4${\footnote{We used the $dP_9$-fibration $Y_4$ instead of a Calabi-Yau 4-fold $Z_4$ precisely for this reason.}}.  This is because the Heterotic 3-fold ${\cal{Z}}_H$ is given by $Z_2=0$ and hence represents a nontrivial divisor class $\pi^*r + p^*t$.  For instance, the restriction of ${\cal{C}}$ to ${\cal{Z}}_H$ is given by
\begin{equation}\begin{split}{\cal{C}}\cdot {\cal{Z}}_H &= \left[5(\sigma + p^*c_1 + \pi^*r)+ p^*(c_1-t)\right]\cdot (\pi^*r+p^*t) \\
&= \left[5\sigma + p^*(6c_1-t)\right]\cdot (\pi^*r+p^*t) \\
&= 5\sigma_H +\pi_H^*(6c_1-t)\,,\end{split}\end{equation}
where $\sigma_H$ is just the restriction of the section $\sigma$ to ${\cal{Z}}_H$ and $\pi_H$ is the elliptic fibration
\begin{equation}\pi_{H} : {\cal{Z}}_H \rightarrow S_2\,.\end{equation}
We recognize this as the Heterotic spectral cover ${\cal{C}}_H$ \cite{Donagi:2004ia}.  We can similarly study the restriction of $\mathbf{10}$ and $\mathbf{\overline{5}}$ "dual matter surfaces" to ${\cal{Z}}_H$
\begin{equation}\begin{split}
\tilde{\Sigma}_{10}\cdot {\cal{Z}}_H &= \left[\sigma_H + \pi_H^*(c_1-t)\right]\cdot_{ {\cal{Z}}_H} {\cal{C}}_H \\
\tilde{\Sigma}_{\mathbf{\overline{5}}}\cdot {\cal{Z}}_H &= 2\sigma_H\cdot_{ {\cal{Z}}_H} \pi_H^*(8c_1-t) + \pi_H^*(3c_1-t)\cdot_{ {\cal{Z}}_H} \pi_H^*(6c_1-t)\,.
\end{split}\end{equation}
We recognize these as the standard matter curves inside ${\cal{C}}_H$ \cite{Donagi:2004ia}.  Finally, the restriction of $\gamma$ to ${\cal{Z}}_H$ is given by
\begin{equation}\gamma\cdot {\cal{Z}}_H = \left[5\sigma_H - \pi_H^*(c_1-t)\right]\cdot {\cal{Z}}_H\,,\end{equation}
which is the standard traceless twist, $\gamma_H$ \cite{Donagi:2004ia}.  It is in fact easy to verify all of these equivalences at the level of equations.

The above restrictions all imply that the Heterotic computation can be phrased directly inside $Y_4$ as
\begin{equation}\begin{split}\chi(R) &= \gamma \cdot_{{\cal{C}}} \tilde{\Sigma}_R\cdot_{ {\cal{C}} } {\cal{Z}}_H|_{ {\cal{C}}} \\
&= {\cal{G}}\cdot \tilde{\Sigma}_R \cdot (\pi^*r + p^*t)\end{split}\label{Hetcounting}\end{equation}
which is almost exactly the form of \eqref{Fcounting}.  That the F-theory and Heterotic computations agree follows from the fact that
\begin{equation}{\cal{G}}\cdot \tilde{\Sigma}_R\cdot p^*t=0\,.\end{equation}
That this works is quite nontrivial and depends crucially on the manner in which $\gamma$ is defined.  In particular, the extra $-5p^*(c_1-t)\cdot \pi^*r$ term in \eqref{Gfluxex} is crucial for this relation to hold.  Because $\pi^*r\cdot {\cal{Z}}_H=0$, this extra piece is invisible to the Heterotic computation but is apparently crucial for the F-theory computation to yield the right result.




\subsection{Chiral Spectrum from "Inherited" $G$-Flux II -- General Case}
\label{subsec:chiralgen}

Let us now turn to $G$-fluxes for more general elliptically fibered Calabi-Yau 4-folds $Z_4$ with base $B_3$ and section $\sigma$ that exhibit $SU(5)$ singularities along a divisor $S_{2}$ inside $B_3$.  We take $z$ to be a holomorphic section on $B_3$ whose vanishing defines $S_{2}$.  In general, $Z_4$ should admit a Weierstrass description
\begin{equation}y^2 = x^3 + fxv^4 + gv^6\,,\end{equation}
where $f$ and $g$ are sections of $K_{B_3}^{-4}$ and $K_{B_3}^{-6}$, respectively.  In general, $f$ and $g$ will be polynomials of holomorphic sections on $B_3$, which can be organized into a series of terms with increasing powers of $z$.  Because $Z_4$ exhibits an $SU(5)$ singularity on $S_2$, a suitable shift of variables should lead to a geometry of the form
\begin{equation}y^2 = x^3 + v\left[b_0 (zv)^5 + b_2 (zv)^3 x + b_3 (zv)^2 y + b_4 zv x^2 + b_5 xy\right] + \ldots\,,\label{fourfold}\end{equation}
where the $\ldots$ contain terms with a higher power of $z$ at each fixed order in $v$.  In this case, we simply define the spectral divisor as
\begin{equation}{\cal{C}} = b_0 (zv)^5 + b_2 (zv)^3 x + b_3 (zv)^2 y + b_4 zv x^2 + b_5 xy\,.\end{equation}
In the neighborhood of the $SU(5)$ singular locus, ${\cal{C}}$ behaves in exactly the same way as the spectral divisor of the $dP_9$ example in the previous section, meaning that it locally behaves like a union of 5 exceptional lines fibered over $S_2$.  We must of course resolve singularities on $S_2$ but, as this is done primarily in patches containing $S_2$, we can mimic the resolution of the $dP_9$ case.  The intersection data \eqref{localintdata} of ${\cal{C}}$ with the resolved curves is therefore identical to the $dP_9$ case, meaning that we can translate the counting formulae directly from the previous section{\footnote{This crucially relies on the fact that our derivation used only local intersection data, ie intersections of the form $C^2$ and $C\cdot \ell$ for the precise combination of $C$'s that had an effective representative near the singular locus.  Had we needed information about $\ell^2$ or intersections involving any other combination of $C$'s, our results could not be carried over.}}.

More specifically, the homological class of ${\cal{C}}$ is given by{\footnote{Note that for $B_3$ a $\mathbb{P}^1$-fibration over $S_2$, this can be written as ${\cal{C}} = 5\sigma + 6\left(p^*c_1+2\pi^*r + p^*t\right)$, which differs from what we found in \eqref{specdivdef} in cases with a Heterotic dual by a factor of $7(\pi^*r+p^*t)$.  This is because the object we are calling the spectral divisor in that special situation was reducible -- it was multiplied by an overall factor of $Z_2^7$.  Six factors of $Z_2$ come outside from the stable degeneration limit and the other appears explcitly in the $dP_9$-fibration \eqref{Y4def}.  This doesn't affect any aspect of the F-theory computation because $Z_2$ is nonzero everywhere along $S_2$.  It is important, however, for determining the appropriate behavior of the union of lines over $S_2$ near $Z_2=0$, where the Heterotic computation is performed.}}
\begin{equation}{\cal{C}} = 5\sigma + 6\pi^*c_1(B_3)\,.\end{equation}
The $\mathbf{10}$ and $\mathbf{\overline{5}}$ "dual matter surfaces" are determined in Appendix \ref{app:mattersurfaces} as
\begin{equation}\begin{split}
\tilde{\Sigma}_{\mathbf{10}} &=  (\sigma + \pi^*S_2)\cdot \pi^*c_1(B_3) \\
\tilde{\Sigma}_{\mathbf{\overline{5}}} &= {\cal{C}}\cdot {\cal{C}} - [y]\cdot {\cal{C}} - (\sigma + \pi^*S_2)\cdot [b_5 x] - 4\pi^*S_2\cdot [x] \\
&= 2\sigma\cdot \pi^*(8c_1(B_3) - 5S_2) + \pi^*c_1(B_3)\cdot \left[18\pi^*c_1(B_3) - 11\pi^*S_2\right]\,.
\end{split}\end{equation}
The general form of the traceless "universal" flux is easily seen to be
\begin{equation}G = {\cal{G}}\cdot {\cal{C}}\qquad {\cal{G}} = 5\sigma - \pi^*c_1(B_3)\,.\end{equation}

We now turn to the chirality formulae.  For this, we use the fact that
\begin{equation}S_2|_{S_2} = -t\qquad c_1(B_3)|_{S_2} = c_1-t\,,\end{equation}
where, as usual, $c_1$ is shorthand for $c_1(S_2)$ and $-t$ is shorthand for $N_{S_2/B_3}$.  Proceeding to evaluate $\chi(10)$ and $\chi(\mathbf{\overline{5}})$, we find
\begin{equation}\begin{split}\chi(10) &= \tilde{\Sigma}_{\mathbf{10}}\cdot {\cal{G}}\cdot \pi^*S_2 \\
&= -(6c_1-t)\cdot_{S_2}(c_1-t) \\
\chi(\mathbf{\overline{5}}) &= \tilde{\Sigma}_{\mathbf{\overline{5}}} \cdot {\cal{G}} \cdot \pi^*S_2 \\
&= -(6c_1-t)\cdot_{S_2}(c_1-t)\,.
\end{split}\end{equation}
We thus see agreement with the formulae that we would obtain by taking the local geometry near $S_2$, embedding inside a global K3-fibration, and performing the computation on the Heterotic side.  While this is a result that we expected, it is comforting to see it arise directly from a computation inside the F-theory 4-fold, $Z_4$.


\subsection{Quantization of G-flux}
We propose that the quantization of flux  $G=i_{{\cal{C}}*}\gamma$ follows from the
quantization of spectral flux $\gamma$ in terms of line bundle
$L_{\cal{C}}$ on ${\cal{C}}$
\begin{equation}
\label{qu} \gamma=c_1(L_{\cal{C}})+\half r_{\cal{C}}\,.
\end{equation}

Here $r_{\cal{C}}$ is the ramification divisor for the map
$p_{\cal{C}}: {\cal{C}} \mapsto B_3$:
\begin{equation}\label{ram}r_{\cal{C}}=c_1({\cal{C}})-p_{\cal{C}}^*c_1(B_3)\,.
\end{equation}
This proposal is motivated by similar condition in Friedman-Morgan-Witten construction of $SU(n)$ bundles $V$ on smooth elliptically fibered
Calabi-Yau 4-folds. In fact, condition (\ref{qu}) ensures that $c_1(V)=0.$
For (\ref{qu}) to make sense in our case, we should first resolve the $A_4$
singularity in $Z_4$ and then consider the line bundle $L_{\cal{C}}$ on a proper
transform of ${\cal{C}}$. Similarly, it is the proper
transform of ${\cal{C}}$ that enters the definition of the ramification divisor.

Let us check that (\ref{qu}) is compatible with quantization of
$\gamma_H=\gamma \cdot {\mathcal{Z}}_H$ in models with Heterotic dual. The consistency requires
\begin{equation} \label{consis}r_{\cal{C}}\cdot {\cal Z}_H=r_H \quad \text{mod}\quad 2\,.\end{equation}
Now recall
\begin{equation}r_H=-{\cal{C}}_H^2-p_{C_H}^*c_1 \quad \text{mod} \quad 2\,.
\end{equation}
Meanwhile,
\begin{equation} r_{\cal{C}}\cdot {\cal Z}_H=\bigl(c_1(Y_4)-{\cal{C}}\bigr)\cdot {\cal{C}} \cdot {\cal Z}_H
-p_{\cal{C}}^*c_1(B_3)\cdot {\cal Z}_H\,.
\end{equation}
Now we use
\begin{equation}c_1(Y_4)=\pi^*r +p^*t,\qquad c_1(B_3)=2r+ \rho^*(c_1+t)\,,
\end{equation}
as well as ${\cal
Z}_H=\pi^*r+p^*t$ and $r(r+\rho^*t)=0$ to prove the consistency (\ref{consis}).

\section{Application: Connecting Semi-local and Global}

Now that we have a proposal for describing $G$-fluxes in a manner intrinsic to F-theory, we turn to an important application, namely the extension of fluxes constructed in so-called "semi-local" models to global fluxes.  We begin by reviewing the connection between the Heterotic spectral cover and Higgs bundles of the 8-dimensional gauge theory on the GUT divisor with emphasis on prior construction of fluxes in semi-local F-theory GUT models \cite{Donagi:2009ra}.  We describe how these are encoded more generally in the spectral divisor and what is needed to ensure that the semi-local fluxes can extend to global ones of the type that we proposed in this note.  We then turn to a simple example of a semi-local model with a reduced monodromy group leading to a $U(1)$ selection rule in the GUT model \cite{Marsano:2009wr}.  After describing some novel fluxes that arise when the monodromy group is reduced, we provide a method of constructing global realizations of this semi-local model, including the novel fluxes.  This includes computations of the chiral spectra of matter fields that are charged under $SU(5)$, which are shown to agree with the semi-local results.  These global models, which do not admit Heterotic duals, come with an additional feature -- singlet fields that carry charge under the additional $U(1)$ that provides the selection rule.  We make a few brief comments about these fields and present a conjecture for how they should be counted.

\subsection{Semi-Local Models and Spectral Covers}

Semi-local F-theory models for $SU(5)$ GUTs are described by a local K3 fibration over a compact complex surface, $S_2$, along which the fiber exhibits an $SU(5)$ singularity.  Writing this in Tate form, we have
\begin{equation}y^2 = x^3 + a_0 z^5v^5 + a_2 z^3v^3 x + a_3 z^2v^2 y + a_4 zvx^2 + a_5 xy\,,
\label{su5tate}\end{equation}
where $x,y,z$ are sections of $4(c_1-t)$, $6(c_1-t)$, and $-t$ respectively.  As usual, $c_1$ is shorthand for $c_1(S_2)$ and $-t$ is a line bundle that we associate with $N_{S_2/B_3}$.  The objects $a_m$ are sections of the bundles $(6-m)c_1-t$ on $S_2$.

Alternatively, one can view the physics near $S_2$ as arising from an 8-dimensional $E_8$ gauge theory with a nontrivial Higgs bundle.  The equivalence between the Higgs bundle data and that of the local geometry plus "local flux" is very explicit and is described in detail in \cite{Donagi:2009ra}.  The Higgs bundle is specified by a "local" spectral cover ${\cal{C}}_{loc}$ that takes the form of a 5-fold cover of $S_2$ inside the total space of $K_{S_2}$
\begin{equation}
{\cal{C}}_{loc}=a_0 s^5 + a_2 s^3 + a_3 s^2 + a_4 s + a_5\,.
\label{localsc}\end{equation}
Here, $s$ is a section of $-c_1$ and the $a_m$ are sections of $\eta-mc_1$ for some class $\eta$.  Defining the projection map
\begin{equation}p_{C_{loc}}:\quad {\mathcal{C}}_{loc}\rightarrow S_2\,,\end{equation}
we obtain the background Higgs field $\Phi$ by $p_{\mathcal{C}_{loc}*}s$ and the gauge bundle by $p_{\mathcal{C}_{loc}*}L$, where $L$ is a line bundle that has been specified on ${\mathcal{C}}_{loc}$.  As before, the identification $\eta=6c_1-t$ allows for a direct map of the sections in \eqref{su5tate} to those in \eqref{localsc}.

To see how ${\cal{C}}_{loc}$ is connected to the Heterotic spectral cover, ${\cal{C}}_H$, we can embed this local geometry into one that is globally K3-fibered.
In that case, the spectral divisor ${\cal{C}}$ is simply \eqref{specdivdef}
\begin{equation}{\cal{C}}=b_0 (Z_1v)^5 + b_2(Z_1v)^3x + b_3(Z_1v)^2y + b_4(Z_1v)x^2 + b_5 xy\,.\end{equation}
As before, we identify the $a_m$ as restrictions of the $b_m$ to ${\cal{Z}}_H$
\begin{equation}a_m = b_m|_{ {\cal{Z}}_H }\,,\end{equation}
 in which case the Heterotic spectral cover is described by
\begin{equation}{\cal{C}}_H: a_0 z^5 + a_2 z^3 x + a_3 z^2 y + a_4 z x^2 + a_5 xy\,.\end{equation}
Because $x\sim s^{-2}$ and $y\sim s^{-3}$ near the zero section, we see that ${\cal{C}}_{loc}$ is just describing the local geometry of ${\cal{C}}_H$ near the zero section.  Similarly, the $(1,1)$-form $L=\gamma_{loc}+\frac{r}{2}$ is just the "local" behavior of the twist $\gamma_H$ up to the shift $\frac{1}{2}r$, where $r$ is the ramification divisor of the covering ${\cal{C}}_{loc}$.  Because the $\mathbf{10}$ and $\mathbf{\overline{5}}$ matter curves are compact inside ${\cal{C}}_{loc}$, it is reasonable to expect that a computation inside ${\cal{C}}_{loc}$ will suffice for studying the spectrum there.  This was beautifully demonstrated in \cite{Donagi:2009ra}, thereby completing the connection between the Higgs bundle and Heterotic pictures.

In semi-local models, it is often important to study spectral covers ${\cal{C}}_{loc}$ which are very nongeneric in the sense that they factor into multiple components.  To see why, recall that the individual sheets of ${\cal{C}}_{loc}$ correspond locally to eigenvalues of $\Phi$ so that their interconnectedness captures the monodromy group associated to the Higgs bundle.  In general, the monodromy group acts on the $SU(5)_{\perp}$ commutant of $SU(5)_{\rm GUT}$ inside $E_8$ as the full Weyl group $S_5$, effectively projecting out any $U(1)\subset SU(5)_{\perp}$ factors that remain from the explicit breaking $E_8\rightarrow SU(5)_{\rm GUT}$.  When ${\cal{C}}_{loc}$ factors, though, the monodromy group is reduced and some $U(1)$ factors remain.  These are expected, from the gauge theory perspective, to introduce selection rules into the superpotential that can solve several phenomenological problems \cite{Tatar:2009jk,Marsano:2009gv,Marsano:2009wr}.
Further, such a factorization generically gives rise to a variety of new supersymmetric fluxes $\gamma_{local}$, which have the interpretation of $U(1)$ fluxes, that give greater flexibility when trying to build 3-generation models \cite{Marsano:2009wr}{\footnote{In addition to this, ${\cal{C}}_{loc}$ can be tuned even further to introduce additional supersymmetric fluxes \cite{Donagi:2009ra,Marsano:2009gv,Marsano:2009wr}.  For simplicity, we will not discuss such tunings in this note.}}.

Factorization of ${\cal{C}}_{loc}$, however, typically does not imply factorization of the corresponding ${\cal{C}}_H$.  When this happens, the novel fluxes in ${\cal{C}}_{loc}$ do not extend to well-defined twists $\gamma_H$ of ${\cal{C}}_H$.  This means that, when a semi-local model of this type is embedded into a global K3 fibration with Heterotic dual, the local flux corresponding to $\gamma_{loc}$ cannot be globally extended{\footnote{If one could cook up an example where ${\cal{C}}_H$ also factors appropriately without also increasing the rank of the GUT gauge group, this would provide an example in which the flux can indeed be extended in a compactification with Heterotic dual.}}.

We cannot use the Heterotic spectral cover to study the extension of local fluxes determined by $\gamma_{loc}$ in more general compactifications.  Fortunately, the spectral divisor ${\cal{C}}$ provides us with a new tool that allows us to do this.  In particular, the sturcture of ${\cal{C}}_{loc}$ can be recovered by studying the geometry of ${\cal{C}}$ near $S_2$.  We do this by taking $z,v\rightarrow 0$ and restricting the sections $b_m$ on $B_3$ to the corresponding sections $a_m$ on $S_2$.  We note that ${\cal{C}}$ depends on $z$ and $v$ through the combination $zv$, which is a section of $\sigma+\pi^*S_2$.  Near $S_2$, $zv$ behaves as a section of the bundle $(\sigma+\pi^*S_2)|_{S_2}=-c_1$, leading to a natural identification of this combination with $s$.  The problem of determining whether a given $\gamma_{loc}$ extends is equivalent to looking for a twist $\gamma$ on ${\cal{C}}$ that reduces to $\gamma_{loc}$ in this limit.

In what follows, we study this issue in the context of a particular example in which ${\cal{C}}_{loc}$ factors into quadratic and cubic pieces \cite{Marsano:2009wr}.  We will review the structure of matter curves, fluxes, and spectra in this setting and then describe a way to construct a global completion in which the spectral divisor ${\cal{C}}$ also factors into two components.  We will then demonstrate that the "dual matter surfaces", fluxes, and spectra map between the semi-local and global constructions in the expected way.



\subsection{Semi-local Models with 3+2 Factorization of ${\cal{C}}_{loc}$}\label{subsec:semilocal32}
\subsubsection{Matter Curves in ${\cal{C}}_{loc}$}\label{subsubsec:Clocmc}

The construction of phenomenologically sound $SU(5)$ GUT models in F-theory requires not only the presence of an $SU(5)$ gauge group and ${\bf 5}_M$ and ${\bf 10}_M$ matter, Higgs fields and Yukawa couplings, but also the absence of dangerous proton decay and R-parity violating operators. Additional gauged $U(1)$ symmetries are particularly useful in realizing these properties. There is two-parameter family of $U(1)$s that are compatible with the MSSM couplings: $U(1)_{PQ}$ and $U(1)_{\chi}$
\begin{equation}\begin{array}{c|c|c}
\text{Field}	& U(1)_{PQ}	 & U(1)_{\chi}    \\ \hline
\mathbf{10}_M 	&			+3	& -1\\
\mathbf{\overline{5}}_M& 		-4		&+ 3\\
\mathbf{5}_H & -6		& +2\\
\mathbf{\overline{5}}_H &   +1		& - 2
\end{array}\end{equation}
From the point of view of the semi-local spectral cover this means that the monodromy group is reduced to a proper subgroup of $S_5$= Weyl$(SU(5)_{\perp})$. Whenever there are additional such $U(1)$ symmetries, these will give rise to further selection rules.

For definiteness we consider as in \cite{Marsano:2009wr} the case of an $S_3 \times \mathbb{Z}_2$ monodromy group. Accordingly, there are two orbits of the fundamental weights of $SU(5)_\perp$, which we will denote by $\lambda_{i=1,2,3}$ and $\lambda_{a=4,5}$. The reduced monodromy corresponds to a $3+2$ factorization of the spectral cover $\mathcal{C}_{loc}$ as in (\ref{localsc})
\begin{equation}
\mathcal{C}_{loc} = \mathcal{C}^{(3)}_{loc} \mathcal{C}^{(2)}_{loc} =  (d_0 s^3 + d_1 s^2 + d_2 s + d_3  ) (e_0 s^2 + e_1 s + e_2 ) =0 \,.
\label{Clocfact}\end{equation}
To ensure that the coefficient  $a_1 = d_1 e_0+ d_0 e_1 =0$, we choose $d_0 = \alpha e_0$ and $d_1 = - \alpha e_1$.
The sections $a_n$ in the local spectral cover are related to the sections in the factored cover as follows
\begin {equation}\label{Clocfactsec}
\begin{aligned}
a_5 &= d_3 e_2 \cr
a_4 &= d_3 e_1+d_2 e_2 \cr
a_3 &= d_3 e_0+d_2 e_1+d_1 e_2\cr
a_2 &=d_2 e_0+d_1 e_1+d_0 e_2 \cr
a_0 &= d_0 e_0 \,.
\end{aligned}
\end{equation}
We further define projections for each component
\begin{equation}p_{n,loc}:{\cal{C}}^{(n)}_{loc}\rightarrow S_2\end{equation}

For practical purposes, it is useful to compactify the canonical line bundle over $S_2$ to the projective bundle $\mathbb{P}({\cal{O}}\oplus K_{S_2})$ \cite{Donagi:2009ra}.  This bundle has two sections, $U$ and $V$, that restrict to homogeneous coordinates on the fiber.  The object $s$ is then an affine coordinate near $U=0$, $s=U/V$.  The bundle of which $U$ is a section is denoted by $\sigma_{loc}$ and the projection down to $S_2$ by $\pi_{loc}$
\begin{equation}\pi_{loc}: \mathbb{P}({\cal{O}}\oplus K_{S_2})\rightarrow S_2\end{equation}
With this notation, $V$ is a section of $\sigma_{loc}+\pi_{loc}^*c_1$, $\sigma_{loc}$ satisfies the identity $\sigma_{loc}(\sigma_{loc}+\pi_{loc}^*c_1)=0$, and the factors of the projectivized ${\cal{C}}_{loc}$ are in the divisor classes
\begin{equation}
\begin{aligned}
\mathcal{C}^{(3)}_{loc}: &\quad  3\sigma_{loc} + \pi_{loc}^*(\eta - 2 c_1  -\xi )   \cr
\mathcal{C}^{(2)}_{loc}: &\quad   2\sigma_{loc} + \pi_{loc}^* (2 c_1 + \xi )   \,,
\end{aligned}
\end{equation}
where $\xi\in H_2(S_2, \mathbb{Z})$ is only constrained by the requirement that the sections remain holomorphic.

The various matter and Yukawa couplings correspond to codimension one and two loci of further symmetry enhancement.
The ${\bf 10}$ matter curve has a component in each factor of the spectral cover given by $\mathcal{C}^{(i)} \cdot \sigma_{loc}$. The ${\bf \overline{5}}$ matter curve arises from the intersection of $\mathcal{C}\cap \tau \mathcal{C}$ \cite{Marsano:2009gv}, where $\tau: \ y \rightarrow -y$.
The various matter loci can be summarized as follows
\begin{equation}\begin{array}{c|c|c|c}
\text{Matter} & \text{Spectral Cover Origin} & \text{Weights} & U(1)_{PQ}\text{ Charge} \\ \hline
\mathbf{10}_{ex}=\mathbf{10}^{(3)} & {\mathcal{C}}^{(3)}_{loc} & \lambda_i & -2 \\
\mathbf{10}_M= \mathbf{10}^{(2)} & {\mathcal{C}}^{(2)}_{loc} & \lambda_a & +3 \\
\mathbf{\overline{5}}_M = \mathbf{\overline{5}}^{(3)} & {\mathcal{C}}^{(3)}_{loc}-{\mathcal{C}}^{(3)}_{loc} & \lambda_i+\lambda_j & -4 \\
\mathbf{5}_H=\mathbf{\overline{5}}^{(2)} & {\mathcal{C}}^{(2)}_{loc}-{\mathcal{C}}^{(2)}_{loc} & \lambda_a+\lambda_b & +6 \\
\mathbf{\overline{5}}_H=\mathbf{\overline{5}}^{(3)(2)} & {\mathcal{C}}^{(3)}_{loc}-{\mathcal{C}}^{(2)}_{loc} & \lambda_i+\lambda_a & +1\end{array}\end{equation}
This assignment of matter curves to GUT multiplets gives rise to a semi-local model that has one additional $U(1)_{PQ}$ symmetry \cite{Marsano:2009wr}{\footnote{Several reasons for implementing $U(1)_{PQ}$ symmetries in F-theory models have been discussed previously in \cite{Marsano:2008jq,Heckman:2008qt}.}}.

In more detail, the loci of matter curves in the spectral cover $\mathcal{C}_{loc}$ can be characterized as follows:
The ${\bf 10}$ matter curve is characterized by
\begin{equation}
\Sigma_{\bf 10}:\qquad a_5= d_3 e_2 =0 \,,
\end{equation}
where the two components correspond to the two orbits of the fundamental representation of $SU(5)_\perp$. The ${\bf \bar{5}}$ matter locus, corresponding to an $SU(6)$ enhancement is given by
\begin{equation}
\Sigma_{\bf \overline{5}}:\qquad P = P_1 P_2 P_3= e_1 (d_3 e_0 + d_2 e_1) \left(d_3 e_1 \left(d_2-e_2 \alpha \right)+e_2 \left(d_2-e_2 \alpha \right){}^2+d_3^2 e_0\right) =0 \,.
\end{equation}
While these describe the loci inside $S_2$ along which matter fields localize, they more properly live on curves inside ${\cal{C}}_{loc}$.  The classes of these matter curves can be identified as follows \cite{Marsano:2009wr}, where we recall that $\eta=6c_1-t$
\begin{equation}\begin{array}{c|c|c}
\text{Curve} & \text{Origin} & \text{Class} \\ \hline
\mathbf{10}^{(2)} & \mathcal{C}_{loc}^{(2)} - \mathcal{C}_{loc}^{(2)} & \sigma_{loc}\cdot\pi_{loc}^{\ast}\xi  \\
\mathbf{10}^{(3)} & \mathcal{C}_{loc}^{(3)}- \mathcal{C}_{loc}^{(3)} & \sigma_{loc}\cdot\pi_{loc}^{\ast}(\eta-5c_1-\xi)  \\
\mathbf{\overline{5}}^{(2)} & \mathcal{C}_{loc}^{(2)}-\mathcal{C}_{loc}^{(2)} & \left[2\sigma_{loc} + \pi_{loc}^{\ast}(2c_1+\xi)\right]\cdot\pi_{loc}^{\ast}(c_1+\xi) \\
\mathbf{\overline{5}}^{(3)(2)} & \mathcal{C}_{loc}^{(3)}- \mathcal{C}_{loc}^{(2)} & 2\left[\sigma_{loc}\cdot\pi_{loc}^{\ast}(2\eta-8c_1-\xi)\right.  \\
& & \left. + \pi_{loc}^{\ast}(\eta-4c_1-\xi)\cdot\pi_{loc}^{\ast}(2c_1+\xi)\right]  \\
\mathbf{\overline{5}}^{(3)} & \mathcal{C}_{loc}^{(3)}- \mathcal{C}_{loc}^{(3)}& 2\sigma_{loc}\cdot\pi_{loc}^{\ast}(\eta-3c_1) \\
& & + (\pi_{loc}^{\ast}\eta)^2 + 14(\pi_{loc}^{\ast}c_1)^2 + (\pi_{loc}^{\ast}\xi)^2  \\
& & + 9 \pi_{loc}^{\ast}c_1\cdot\pi_{loc}^{\ast}\xi - 2\pi_{loc}^{\ast}\eta\cdot\pi_{loc}^{\ast}\xi \\
& & - 7\pi_{loc}^{\ast}c_1 \cdot \pi_{loc}^{\ast}\eta \\
\end{array}\end{equation}



\subsubsection{Fluxes in ${\cal{C}}_{loc}$}

For a generic spectral cover ${\cal{C}}_{loc}$, there is only one choice of traceless $\gamma_{loc}$ that can be introduced \cite{Donagi:2009ra}
\begin{equation}\gamma_{loc} = 5({\cal{C}}_{loc}\cdot \sigma_{loc}) - p_{C_{loc}}^*p_{C_{loc}*} ({\cal{C}}_{loc}\cdot \sigma_{loc})\end{equation}
When ${\cal{C}}_{loc}$ factors, however, several new types of fluxes can be turned on{\footnote{Even more "nonuniversal" fluxes can be turned on if we tune ${\cal{C}}_{loc}$ further.  For simplicity, we do not discuss these but the generalization is straightforward.}} \cite{Marsano:2009wr}.  For now, we review three such fluxes and then discuss their "extension" to well-defined $G$-fluxes using the spectral divisor.

When ${\cal{C}}_{loc}$ factors, we must specify fluxes $\gamma_{n,loc}$ on each component ${\mathcal{C}}^{(n)}_{loc}$.  These do not have to be individually traceless but rather need only satisfy a net traceless condition
\begin{equation}p_{3,loc\,*}\gamma_{3,loc} + p_{2,loc\,*}\gamma_{2,loc}=0\label{nettraceless}\end{equation}
Building blocks for $\gamma_{3,loc}$ and $\gamma_{2,loc}$ that exist for a generic 3+2 factored ${\cal{C}}_{loc}$ include intersections with $\sigma_{loc}$
\begin{equation}
\sigma_{loc} \cdot \mathcal{C}_{loc}^{(n)}\,,
\end{equation}
or pullbacks of curves $\Sigma$ in $S_2$
\begin{equation}
p_{n,loc}^* \Sigma = \pi_{loc}^*\Sigma \cdot \mathcal{C}_{loc}^{(n)} \,.
\end{equation}
We can form linear combinations that are individually traceless on each ${\cal{C}}_{loc}^{(n)}$ by
\begin{equation}
\tilde{\gamma}_{loc,n} = n (\sigma_{loc}\cdot {\cal{C}}_{loc}^{(n)})- p_n^* p_{n*} (\sigma\cdot {\cal{C}}_{loc}^{(n)}) \,.
\end{equation}
In the 3+2-factored case, we get two such fluxes which can be written as intersections of divisors in $\mathbb{P}({\cal{O}}\oplus K_{S_2})$ with components of the local spectral cover as
\begin{equation}\tilde{\gamma}_{3,loc} = {\cal{C}}^{(3)}_{loc}\cdot \left[3\sigma_{loc} - \pi_{loc}^{\ast}(\eta-5c_1-\xi)\right]\,,\qquad
\tilde{\gamma}_{2,loc} = {\cal{C}}_{loc}^{(2)}\cdot \left[2\sigma_{loc} - \pi_{loc}^{\ast}\xi\right]\,.\end{equation}
In \cite{Marsano:2009wr} we further considered fluxes of the type
\begin{equation}
\begin{aligned}
\tilde{\rho}_{loc} &= 2p_3^*\rho - 3p_2^*\rho \,,
\end{aligned}
\end{equation}
where $\rho$ is a curve class in $S_2$.  This flux is not individually traceless on ${\mathcal{C}}^{(3)}_{loc}$ or ${\mathcal{C}}^{(2)}_{loc}$ but satisfies the net traceless constraint \eqref{nettraceless},  where $\rho$ is some curve class in $S_2$.  There are other traceless combinations that one can form \cite{Marsano:2009wr} but, to keep things simple, we focus on these three.

The intersection table of the ${\tilde \gamma}_{loc}$-fluxes above with the matter curves is \cite{Marsano:2009wr}
\begin{equation}\begin{array}{c|c|c|c}\text{Curve} & \text{Class} & \tilde{\gamma}_{3,loc} & \tilde{\gamma}_{2,loc} \\ \hline
\mathbf{10}^{(2)} & \sigma_{loc}\cdot\pi_{loc}^{\ast}\xi & 0 & - \xi\cdot_{S_2}(2c_1+\xi) \\
\mathbf{10}^{(3)}  & \sigma_{loc}\cdot \pi_{loc}^{\ast}(\eta-5c_1-\xi) & -(\eta-2c_1-\xi)\cdot_{S_2} (\eta-5c_1-\xi) & 0 \\
\mathbf{\overline{5}}^{(2)} & [2\sigma_{loc} + \pi_{loc}^{\ast}(2c_1+\xi)]\cdot \pi^{\ast}(c_1+\xi)& 0 & 0 \\
\mathbf{\overline{5}}^{(3)(2)}  & 2\left[\sigma_{loc}\cdot\pi_{loc}^{\ast}(2\eta-8c_1-\xi)\right. &-2(\eta-4c_1-2\xi)\cdot_{S_2} (\eta-5c_1-\xi)& -\xi\cdot_{S_2} (2c_1+\xi)\\
&   \left. +\pi_{loc}^{\ast}(\eta-4c_1-\xi)\cdot \pi_{loc}^{\ast}(2c_1+\xi)\right] & & \\
\mathbf{\overline{5}}^{(3)}  & 2\sigma_{loc}\cdot \pi_{loc}^{\ast}(\eta-3c_1) & (\eta-5c_1-\xi)\cdot_{S_2}(\eta-6_1-3\xi) & 0 \\
&  + \pi_{loc}^{\ast}(\eta)^2 + 14\pi_{loc}^{\ast}c_1^2 + \pi_{loc}^{\ast}\xi^2 & & \\
&  +9\pi_{loc}^{\ast}c_1\cdot\pi_{loc}^{\ast}\xi - 2\pi_{loc}^{\ast}\eta\cdot \pi_{loc}^{\ast}\xi & & \\
&   - 7\pi_{loc}^{\ast}c_1\cdot \pi_{loc}^{\ast}\eta & &
\end{array}\end{equation}
Simillarly, the intersections of $\tilde{\rho}_{loc}$ with the matter curves are
\begin{equation}\begin{array}{c|c|c}
\text{Curve}  & \text{Class} & \tilde{\rho}_{loc} \\ \hline
\mathbf{10}^{(2)}  & \sigma_{loc}\cdot\pi_{loc}^{\ast}\xi & -3\rho\cdot_{S_2} \xi \\
\mathbf{10}^{(3)}  & \sigma_{loc}\cdot\pi_{loc}^{\ast}(\eta-5c_1-\xi) & 2\rho\cdot_{S_2} (\eta-5c_1-\xi) \\
\mathbf{\overline{5}}^{(2)}  & \left[2\sigma_{loc} + \pi_{loc}^{\ast}(2c_1+\xi)\right]\cdot\pi^{\ast}(c_1+\xi) & -6\rho\cdot_{S_2} (c_1+\xi) \\
\mathbf{\overline{5}}^{(3)(2)}  & 2\left[\sigma_{loc}\cdot\pi_{loc}^{\ast}(2\eta-8c_1-\xi)\right. & -\rho\cdot_{S_2} (2\eta-8c_1-\xi) \\
& \left. + \pi_{loc}^{\ast}(\eta-4c_1-\xi)\cdot\pi_{loc}^{\ast}(2c_1+\xi)\right] & \\
\mathbf{\overline{5}}^{(3)} & 2\sigma_{loc}\cdot\pi_{loc}^{\ast}(\eta-3c_1) & 4\rho\cdot_{S_2} (\eta-3c_1) \\
& + (\pi_{loc}^{\ast}\eta)^2 + 14(\pi_{loc}^{\ast}c_1)^2 + (\pi_{loc}^{\ast}\xi)^2 & \\
 & + 9 \pi_{loc}^{\ast}c_1\cdot\pi_{loc}^{\ast}\xi - 2\pi_{loc}^{\ast}\eta\cdot\pi_{loc}^{\ast}\xi & \\
 & - 7\pi_{loc}^{\ast}c_1 \cdot \pi_{loc}^{\ast}\eta \\
\end{array}\end{equation}

\subsection{Global Embedding of 3+2 Model}
\label{subsec:global32}

We now discuss one way to construct a global completion of the semi-local model of section \ref{subsec:semilocal32}.  Starting with a local spectral cover \eqref{Clocfact} with sections $a_m$ of the form \eqref{Clocfactsec}, we first choose sections $\tilde{d}_m$, $\tilde{e}_n$, and $\tilde{\alpha}$ on $B_3$ of the bundles{\footnote{Of course, we assume that all bundles admit holomorphic sections.}}
\begin{equation}\begin{array}{c|c}\text{Section} & \text{Bundle} \\ \hline
\tilde{d}_m & (4-m)c_1(B_3)-(3-m)S_2-\hat{\xi} \\
\tilde{e}_n & (2-m)c_1(B_3)-(2-m)S_2+\hat{\xi}\\
\tilde{\alpha} & 2c_1(B_3)-S_2-2\hat{\xi}
\end{array}\end{equation}
where $\hat{\xi}$ is a bundle on $B_3$ whose restriction to $S_2$ is $\xi$
\begin{equation}\hat{\xi}|_{S_2} = \xi\,.
\end{equation}
From these, we construct
\begin{equation}\begin{split}
b_5 &= \tilde{d}_3\tilde{e}_2 \\
b_4 &= \tilde{d}_3\tilde{e}_1 + \tilde{d}_2\tilde{e}_2 \\
b_3 &= \tilde{d}_3\tilde{e}_0 + \tilde{d}_2\tilde{e}_1 + \tilde{d}_1\tilde{e}_2 \\
b_2 &= \tilde{d}_2\tilde{e}_0 + \tilde{d}_1\tilde{e}_1 + \tilde{d}_0\tilde{e}_2 \\
b_0 &= \tilde{d}_0\tilde{e}_0
\label{factbs}\end{split}\end{equation}
where
\begin{equation}\tilde{d}_0 = \tilde{\alpha}\tilde{e}_0\qquad \tilde{d}_1 = -\tilde{\alpha}\tilde{e}_1\,,\end{equation}
in order to ensure that
\begin{equation}b_1 = \tilde{d}_1\tilde{e}_0 + \tilde{d}_0 \tilde{e}_1 = 0\,.\end{equation}
We then construct a 4-fold $Z_4$ as
\begin{equation}y^2 = x^3 + {\cal{C}}\,,\label{truncfourfold}\end{equation}
where ${\cal{C}}$ is the equation that we use to define the spectral divisor
\begin{equation}{\cal{C}} = b_0 (zv)^5 + b_2 (zv)^3 x + b_3 (zv)^2 y + b_4 (zv) x^2 + b_5 xy\,.\label{specdivagain}\end{equation}
Note that \eqref{truncfourfold} is precisely as in \eqref{fourfold} except that the series in $z$ is explicitly truncated as in the constructions of \cite{Marsano:2009gv,Marsano:2009wr}.  What this truncation buys us is a global factorization of the spectral divisor, ${\cal{C}}$.  In the presence of higher order terms as in \eqref{fourfold}, \eqref{specdivagain} will not factor in general, even though it appears to near the zero section were $x\sim s^{-2}$ and $y\sim s^{-3}$.  When the series is truncated, though, $y^2=x^3$ along the entire intersection of ${\cal{C}}$ with $Z_4$.  This means that, on ${\cal{C}}$, we can write{\footnote{More specifically, $\zeta=y/x$, which is a meromorphic section of $\sigma+\pi^*c_1(B_3)$ on $Z_4$.  The restriction of $\zeta$ to ${\cal{C}}$ for the particular four-fold \eqref{truncfourfold} is holomorphic.}}.
\begin{equation}y=\zeta^3\qquad x=\zeta^2\,,\label{xyzeta}\end{equation}


Because of \eqref{xyzeta}, we see that inside the four-fold defined by the "truncated" equation \eqref{truncfourfold}, ${\cal{C}}$ factors into two components when the $b_m$ are specified as in \eqref{factbs}.
\begin{equation}{\cal{C}}\rightarrow \left(\tilde{d}_0(zv)^3 + \tilde{d}_1 (zv)^2\zeta + \tilde{d}_2 (zv)\zeta^2 + \tilde{d}_3\zeta^3\right)\left(\tilde{e}_0(zv)^2 + \tilde{e}_1(zv)\zeta + \tilde{e}_2\zeta^2\right)\,.\end{equation}
We will call these componens ${\cal{C}}^{(3)}$ and ${\cal{C}}^{(2)}$, respectively.  As divisors inside $Z_4$, they are in the classes
\begin{equation}{\cal{C}}^{(3)} = 3\sigma + \pi^*(4c_1(B_3)-\hat{\xi})\qquad {\cal{C}}^{(2)} = 2\sigma + \pi^*(2c_1(B_3)+\hat{\xi})\,.\end{equation}

The local factorization of ${\cal{C}}_{loc}$ indicated that what looks near $S_2$ to be exceptional lines of $\ell^{(i)}$ fall into two distinct groups, of three and two lines respectively, that are not mixed by monodromy.   The object ${\cal{C}}^{(3)}$ (${\cal{C}}^{(2)}$) behaves, near $S_2$, as a union of the three (two) lines of the first (second) group fibered over $S_2$.   The lines $\ell^{(i)}$ have different intersection numbers with the curves $C^{(a)}$ that degenerate when the singularity type on $S_2$ enhances, though.  This means that ${\cal{C}}^{(3)}$, for instance, will have different intersection numbers with the two curves $C^{(a)}$'s that degenerate on the two distinct $\mathbf{10}$ matter curves.  This means that the two $C^{(a)}$'s are homologically distinct inside $Z_4$ and the global monodromy group is reduced by the factorization just as the local monodromy group was.
One therefore expects that an additional $U(1)$ symmetry, along with its selection rules, is preserved in the 4-dimensional theory{\footnote{This will come with a $U(1)$ gauge boson that is typically anomalous.  We expect the gauge boson to be lifted but do not immediately know the mechanism responsible for it.  This would be interesting to clarify.}}.

With a globally factored spectral divisor, we can now construct global extensions of the fluxes $\tilde{\gamma}_{3,loc}$, $\tilde{\gamma}_{2,loc}$, and $\tilde{\rho}_{loc}$ of the previous section.  In particular, we define
\begin{equation}\begin{split}\tilde{\gamma}_3 &= 3({\cal{C}}^{(3)}\cdot \sigma)-p_3^*p_{3*}({\cal{C}}^{(3)}\cdot\sigma)\\
&= {\cal{C}}^{(3)}\cdot \left[3\sigma - \pi^*(c_1(B_3)-\hat{\xi})\right] \\
\tilde{\gamma}_2 &= 2({\cal{C}}^{(2)}\cdot\sigma) - p_2^*p_{2*}({\cal{C}}^{(2)}\cdot\sigma)\\
&={\cal{C}}^{(2)} \cdot \left[2\sigma - \pi^*\hat{\xi}\right] \\
\tilde{\rho} &= 2{\cal{C}}^{(3)}\cdot \pi^*\hat{\rho} - 3{\cal{C}}^{(2)}\cdot \pi^*\hat{\rho}\,.
\end{split}\end{equation}
Here, $\hat{\rho}$ is a divisor inside $B_3$.  In what follows, we will often refer to the object $\rho$, its restriction to $S_2$
\begin{equation}\rho = \hat{\rho}|_{S_2}\end{equation}

We now turn to a computation of the chiral spectrum.  For this, we need the "dual matter surfaces" corresponding to the various matter curves of section \ref{subsubsec:Clocmc}.  As described in Appendix \ref{app:mattersurfaces}, we find
\begin{equation}\begin{array}{c|c}
\text{Field} &  \text{Dual Matter Surface} \\ \hline
\mathbf{10}^{(3)}  & (\sigma + \pi^*S_2)\cdot \pi^*(c_1(B_3)-\hat{\xi}) \\
\mathbf{10}^{(2)}  & (\sigma + \pi^*S_2)\cdot \pi^*\hat{\xi} \\
\mathbf{\overline{5}}^{(3)}  & 2\sigma\cdot \pi^*(3c_1(B_3)-2S_2)+8(\pi^*c_1(B_3))^2 + \pi^*\hat{\xi}\cdot \pi^*(S_2+\hat{\xi}) - \pi^*c_1(B_3)\cdot \pi^*(5S_2+3\hat{\xi}) \\
\mathbf{\overline{5}}^{(2)} & 2\sigma\cdot \pi^*(c_1(B_3)-S_2+\hat{\xi}) + \pi^*(2c_1(B_3) + \hat{\xi})\cdot \pi^*(c_1(B_3)-S_2+\hat{\xi})\\
\mathbf{\overline{5}}^{(3)(2)}& 2\left[\sigma\cdot \pi^*(4c_1(B_3)-2S_2-\hat{\xi}) + 4(\pi^*c_1(B_3))^2 - (\pi^*\hat{\xi})^2 - 2\pi^*c_1(B_3)\cdot\pi_*S_2\right]
\end{array}\end{equation}

Because the fluxes all take the form of an intersections of a divisor ${\cal{G}}$ in $Z_4$ with a component of ${\cal{C}}$, we can compute their induced chiralities as intersections in $Z_4$.  More specifically, the general chirality formula \eqref{Ftheoryform} becomes
\begin{equation}\chi(R) = \pi^*S_2\cdot {\cal{G}}\cdot \hat{\Sigma}_R\,,\end{equation}
where $\hat{\Sigma}_R$ is the corresponding "dual matter surface".  Note that ${\cal{G}}$ breaks up into pieces ${\cal{G}}^{(3)}$ and ${\cal{G}}^{(2)}$ depending on the component of ${\cal{C}}$ in which it lives.  For this reason, we must keep track of which component(s) are relevant for the computation on a given "dual matter surface".  The net chirality induced by one unit of $\tilde{\gamma}_3$, $\tilde{\gamma}_2$, or $\tilde{\rho}$ flux is listed below.  As expected, we have complete agreement with results of the semi-local picture.
\begin{equation}\begin{array}{c|c|c|c}
\text{Field} & \tilde{\gamma}_3 & \tilde{\gamma}_2 & \tilde{\rho} \\ \hline
\mathbf{10}^{(3)} & -(c_1-t)\cdot_{S_2} (4c_1-t-\xi) & 0 & 2(c_1-t-\xi)\cdot_{S_2}\rho \\
\mathbf{10}^{(2)} & 0 & -(2c_1+\xi)\cdot_{S_2}\xi & -3\xi\cdot_{S_2}\rho \\
\mathbf{\overline{5}}^{(3)} & -(c_1-t-\xi)\cdot_{S_2}(t+3\xi) & 0 & 4(3c_1-t)\cdot_{S_2}\rho \\
\mathbf{\overline{5}}^{(2)} & 0 & 0 & -6(c_1+\xi)\cdot_{S_2}\rho \\
\mathbf{\overline{5}}^{(3)(2)} & -2(c_1-t-\xi)\cdot_{S_2}(2c_1-t-2\xi) & -\xi\cdot_{S_2}(2c_1+\xi) & -(4c_1-2t-\xi)\cdot_{S_2}\rho
\end{array}\end{equation}
Note that one has to take some care when computing chiralities of $\mathbf{\overline{5}}^{(3)(2)}$.  A flux on ${\cal{C}}^{(3)}$ only intersects that part of the $\mathbf{\overline{5}}^{(3)(2)}$ dual matter surface that sits inside ${\cal{C}}^{(3)}$.  The class of this part of the dual matter surface is $\frac{1}{2}$ that of the total $\mathbf{\overline{5}}^{(3)(2)}$ dual matter surface.

\subsection{Charged Singlets in the 3+2 Model}

When the spectral divisor factors, we expect that a $U(1)\subset SU(5)_{\perp}\subset E_8$ is globally preserved.  It follows, then, that some of the $SU(5)_{\rm GUT}$ singlets that descend from the adjoint of $E_8$ will carry $U(1)$ charge.  These singlets should localize on a curve of $SU(2)$ enhancement that meets $S_2$ at isolated points where the singularity type is $SU(7)$.  Some aspects of the singlet locus were already studied in \cite{Marsano:2009wr} in the context of the specific 3+2 factorization of section \ref{subsubsec:Clocmc}.  In the semi-local picture, the singlets are identified with vanishing of the combinations $\lambda_i-\lambda_j$.  From the relation of $a_m$ to symmetric polynomials in the $\lambda_i$, one can motivate{\footnote{Actually, by $F_3$, we mean a particular combination of $b_m$'s whose restriction to $S_2$ is the appropriate combination of $a_m$'s.}} a combination $F_3$, which is basically $\prod_{i=1}^3\prod_{a=4}^{5}(\lambda_i-\lambda_a)^2$, that should be one of the defining equations of the singlet locus.  It was demonstrated in \cite{Marsano:2009wr} that this is indeed the case.  One would also like to associate a curve inside ${\cal{C}}_{loc}$ to the charged singlets and the natural guess is that this is just ${\cal{C}}_{loc}^{(3)}\cap {\cal{C}}_{loc}^{(2)}$.  In \cite{Marsano:2009wr}, it was shown that the projection of this curve down to $S_2$ is nothing other than $F_3$.

One problem with the identification of a singlet curve inside ${\cal{C}}_{loc}$ is that it intersects the "divisor at infinity".  In other words, it is a noncompact curve inside $K_{S_2}$.  This reflects the fact that singlet fields do not localize on $S_2$ so are not completely captured by local geometry near $S_2$.  Some conjectures were made in \cite{Marsano:2009wr} but a proper treatment requires going beyond ${\cal{C}}_{loc}$.

The spectral divisor ${\cal{C}}$ allows us to give a better description.  Near $S_2$, we recall that ${\cal{C}}$ behaves like a union of exceptional lines fibered over $S_2$.  Singlets should be associated with differences of exceptional lines so a useful notion of "dual matter surface" for singlets should be ${\cal{C}}^{(3)}\cap {\cal{C}}^{(2)}$.  Of course, the singlet locus extends outside of the region near $S_2$ where we know how to associate ${\cal{C}}^{(3)}$ and ${\cal{C}}^{(2)}$ with exceptional lines.  The geometry near the $SU(2)$ singular locus should itself look like a local K3-fibration, though, which can be used to see whether ${\cal{C}}^{(3)}\cap {\cal{C}}^{(2)}$ is really playing the role of a "dual matter surface" as defined in section \ref{subsec:chirform}.  It is hard to study this in detail, because the $SU(2)$ singular locus is very complicated.  In this section, we just peform a simple check, namely to verify that ${\cal{C}}^{(3)}\cap {\cal{C}}^{(2)}$ contains the $SU(2)$ enhancement locus.

To study the singlet  locus, we turn first to the discriminant.  For the geometry \eqref{truncfourfold}, it was shown in \cite{Marsano:2009wr} that the discriminant takes the form
\begin{equation}
\Delta = z^5 P_5 (d_n, e_m, z) \,,
\end{equation}
The singlets localize along the subspace, where the singularity type enhances from $U(1)$ to $SU(2)$. This happens precisely when $P_5$ has a double root, or alternatively, it is captured by the vanishing of the discriminant of $P_5$ with respect to $z$ which takes the form
\begin{equation}
\Delta_P =  b_5^5 F_1 F_2 F_3^2 G^3 \,.
\end{equation}
The constraint that $P_5(z) = P_5'(z) =0$ can be reduced iteratively to the relations
\begin{equation}\label{SingLoc}
\hbox{Singlet Locus:}\qquad F_3=0 \quad \hbox{and}\quad   z= z_* \,,
\end{equation}
where $z_*$ solves a linear equation that is obtained by successively reducing the degree of $P_5$ and $P_5'$ by substitution. Note that $F_3=0$ is precisely the analog of the charged singlet locus in the semi-local case.
Inserting the factored spectral cover form of the sections $b_n$ \eqref{factbs} we find
\begin{equation}
F_3= e_2 \left(2 d_2 \left(e_1^2-e_0 e_2\right) \alpha +d_2^2 e_0+e_2 \left(2 e_1^2+e_0 e_2\right) \alpha
   ^2\right)-d_3 e_1 \left(e_0 \left(d_2-5 e_2 \alpha \right)+2 e_1^2 \alpha \right)+d_3^2 e_0^2 \,.
\end{equation}

The locus of the $SU(2)$ enhancement, i.e. the singlet locus, is obtained by reinserting (\ref{SingLoc}) into the Weierstrass equation.
Schematically this takes the form
\begin{equation}\label{Weier}
y^2 = x^3 + \mathcal{C}(z= z_*)  \,,
\end{equation}
where in $\mathcal{C}$ the relations (\ref{SingLoc}) are used.

To relate this to the proposed singlet "dual matter surface" inside the spectral divisor, i.e. $\mathcal{C}^{(3)} \cdot \mathcal{C}^{(2)} \supset (F_3=0)$ and $y^2 = x^3$, we parametrize $x= \zeta^2$ and $y= \zeta^3$.
Inserting this parametrization as well as (\ref{SingLoc}) into (\ref{Weier}) we obtain
\begin{equation}
\mathcal{C}^{(3)}(z= z_*) \mathcal{C}^{(2)}(z= z_*)  = (\zeta- \zeta_*)^2 Q_3(\zeta) \,.
\end{equation}
Furthermore it can be checked that $\zeta_*$ is part of the intersection $\mathcal{C}^{(3)} \cdot \mathcal{C}^{(2)}$. This means precisely that the singlet locus inside the spectral divisor, i.e. the singlet "dual matter surface", 
covers the singlet matter curve, $F_3 = 0$ and $z = z_*$ along with (\ref{Weier}).

With ${\cal{C}}^{(3)}\cdot {\cal{C}}^{(2)}$ as a candidate dual matter surface for charged singlets, we can conjecture a formula for counting their net chirality with respect to $U(1)$ charge
\begin{equation}
\chi_{\bf 1} = \mathcal{C}^{(3)} \cdot \mathcal{C}^{(2)} \cdot (\Gamma_3- \Gamma_2) \cdot \pi^* F_3 \,.
\end{equation}
where $\Gamma_n$ is the net twist on ${\cal{C}}^{(n)}$.  Verifying this will require a more detailed study of the geometry near the singlet curve.


\section*{Acknowledgements}

We are grateful to T. Watari for several helpful discussions and to Y.~Chung for pointing out a typo in v1.  The work of JM was supported by DOE grant DE-FG02-90ER-40560.  The work of SSN was supported in part by the National Science Foundation under grant No. PHY05-51164. JM and SSN are grateful to the Institute for the Physics and Mathematics of the Universe, the University of Tokyo, and the organizers of the IPMU Workshop on Elliptic Fibrations and F-theory for hospitality.  JM and SSN would also like to thank the Max Planck Institut f\"ur Physik  in Munich and the organizers of the MPI Workshop on GUTs and Strings for their hospitality.  JM would also like to thank the Perimeter Institute for Theoretical Physics and the Michigan Center for Theoretical Physics for hospitality as well as the organizers of the String Vacuum Project 2010 Meeting.  SSN is grateful to the  Caltech theory group for their generous hospitality.  JM and NS  are grateful to the Kavli Institute for Theoretical Physics for hospitality and we would all like to thank the organizers of the KITP workshop "Strings at the LHC and in the Early Universe" for providing a stimulating research environment.

\appendix

\section{The $dP_9$ Fibration}

Let us write down all details of the $dP_9$ fibration in earnest.  We have
\begin{equation}y^2 = x^3 + \tilde{f} x + \tilde{g}\end{equation}
\begin{equation}\tilde{f} = \sum_{m=0}^4 f_m Z_1^m Z_2^{4-m}\qquad \tilde{g} = \sum_{n=0}^6 g_n Z_1^n Z_2^{6-n}\,.\end{equation}
Sections are as follows
\begin{equation}\begin{array}{c|c}\text{Section} & \text{Bundle} \\ \hline
Z_1 & r \\
Z_2 & r+t \\
x & 2\left[c_1(B_3) - (r+t)\right] = 2(c_1+r) \\
y & 3\left[c_1(B_3) - (r+t)\right] = 3(c_1+r) \\
f_m & 4(c_1-t) + mt \\
g_n & 6(c_1-t) + nt\end{array}\end{equation}
The Heterotic 3-fold is given by $Z_2=0$ and hence is in the class $r+t$.

In the 4-fold, the cohomology ring is generated by that of $B_3$ plus an additional class $\sigma$ that defines the section of the fibration.  The class of $B_3$ inside $Y_4$ is simply $\sigma$ and, furthermore, $\sigma$ satisfies the relation
\begin{equation}\sigma\left(\sigma + \frac{1}{3}[y]\right)=0\,.\end{equation}
To see this we realize our 4-fold as a divisor in a $\mathbb{P}^2$-fibration over $B_3$ via the equation
\begin{equation}VY^2 = X^3 + \tilde{f}XV^2 + \tilde{g}V^3\,.\end{equation}
If $V$ is a section of $\alpha$, which is the bundle that restricts to ${\cal{O}}(1)$ on the $\mathbb{P}^2$ fibers, then $X$ is a section of $\alpha + [x]$ and $Y$ is a section of $\alpha + [y]$ where $[x]$ and $[y]$ are the classes of $x$ and $y$ above.  We have the equivalence relation inside the 5-fold
\begin{equation}\alpha\left(\alpha + [x]\right)\left(\alpha + [y]\right)=0\,.\end{equation}
Because the class of our 4-fold is $3(\alpha + [x])$, we have the following equivalence relation on sections in the 4-fold
\begin{equation}\alpha\left(\alpha + [y]\right)=0\,.\end{equation}
Finally, we note that $\alpha=0$ gives 3 times the zero section of the elliptic fibration, $\sigma$, so we are finally left with
\begin{equation}\sigma \left(\sigma + \frac{1}{3}[y]\right)=0\,,\end{equation}
or{\footnote{Note that for an elliptically fibered Calabi-Yau 4-fold, $[y]=3c_1(B_3)$ so that this would lead to the usual relation $\sigma (\sigma + c_1(B_3))=0$.}}
\begin{equation}\sigma\left(\sigma + c_1+r\right)=0\,.\end{equation}


\section{"Dual  Matter Surfaces" for Unfactored and 3+2-Factored ${\cal{C}}$}
\label{app:mattersurfaces}

In this Appendix, we briefly describe the structure of the "dual matter surfaces" used in the main text for both the generic unfactored spectral divisor of section \ref{subsec:chiralgen} and the factored one of section \ref{subsec:global32}.

\subsection{Unfactored Spectral Divisor of Section \ref{subsec:chiralgen}}

We start with a generic spectral divisor
\begin{equation}{\cal{C}} = b_0(zv)^5 + b_2(zv)^3x + b_3(zv)^2y + b_4(zv)x^2 + b_5xy\label{appspecdiv}\end{equation}
inside the 4-fold \eqref{fourfold}.  Recall that objects here are sections of the following bundles
\begin{equation}\begin{array}{c|c}
\text{Section} & \text{Bundle} \\ \hline
z & \pi^*S_2 \\
v & \sigma \\
x & 2(\sigma+\pi^*c_1(B_3)) \\
y & 3(\sigma+\pi^*c_1(B_3)) \\
b_m & (6-m)\pi^*c_1(B_3)-(5-m)\pi^*S_2
\end{array}\end{equation}
It is also helpful to remember that
\begin{equation}\sigma\cdot (\sigma+\pi^*c_1(B_3))=0\end{equation}

The $\mathbf{10}$ dual matter surface is obtained by noting that, when $b_5=0$, \eqref{appspecdiv} factors into two pieces.  This reflects the fact that, when we go near $S_2$, ${\cal{C}}$ looks locally like a union of exceptional lines fibered over $S_2$, one of which breaks off from the rest when $b_5=0$.  The $\mathbf{10}$ dual matter surface is precisely this factor
\begin{equation}(zv)=0\quad b_5=0\end{equation}
which is in the class
\begin{equation}\tilde{\Sigma}_{\mathbf{10}} = (\sigma+\pi^*S_2)\cdot \pi^*c_1(B_3)\end{equation}

The $\mathbf{\overline{5}}$ dual matter surface sits inside the intersection ${\cal{C}}\cap \tau {\cal{C}}$ where $\tau$ is the involution $y\rightarrow -y$ that locally behaves{\footnote{Recall that locally the $\ell^{(i)}$ behave like exceptional lines of a $dP_9$.  Here, $x_9$ refers to the anti-canonical curve of the $dP_9$.}} like $\ell^{(i)}\rightarrow x_9-\ell^{(i)}$.  This intersection is described by the equations
\begin{equation}\begin{split}0&=(zv)\left[b_0(zv)^4 + b_2(zv)^2x + b_4 x^2\right]\\
0&= y\left[b_3(zv)^2 + b_5 x\right]
\end{split}\end{equation}
This intersection has many different components.  It is clear that what we want is the intersection of the two terms in $[\,]$'s less  the locus $x=z=0$, which is just the $SU(5)_{\rm GUT}$ singular locus.  As such, we have
\begin{equation}\begin{split}\tilde{\Sigma}_{\mathbf{\overline{5}}} &= {\cal{C}}\cdot {\cal{C}} - [y]\cdot[b_4x^2] - [zv]\cdot [b_5x] - [y]\cdot[zv]-4[x]\cdot [z]\\
&= 2\sigma\cdot\pi^*(8c_1(B_3)-5S_2)+\pi^*c_1(B_3)\cdot \pi^*\left[18c_1(B_3)-11S_2\right]
\end{split}\end{equation}

\subsection{Factored Spectral Divisor of Section \ref{subsec:global32}}
We now turn to the factored spectral divisor
\begin{equation}{\cal{C}}={\cal{C}}^{(3)}{\cal{C}}^{(2)}\end{equation}
with
\begin{equation}\begin{split}{\cal{C}}^{(3)} &= \tilde{\alpha}\tilde{e}_0(zv)^3 -\tilde{\alpha} \tilde{e}_1(zv)^2\zeta + \tilde{d}_2(zv)\zeta^2 + \tilde{d}_3\zeta^3 \\
{\cal{C}}^{(2)} &= \tilde{e}_0(zv)^2 + \tilde{e}_1 (zv)\zeta + \tilde{e}_2\zeta^2
\end{split}\end{equation}
where the precise notion of what we mean by $\zeta$ is discussed below \eqref{xyzeta}.  We recall that the objects above are sections of the bundles
\begin{equation}\begin{array}{c|c}\text{Section} & \text{Bundle} \\ \hline
\tilde{d}_m & (4-m)c_1(B_3)-(3-m)S_2-\hat{\xi} \\
\tilde{e}_n & (2-m)c_1(B_3)-(2-m)S_2+\hat{\xi}\\
\tilde{\alpha} & 2c_1(B_3)-S_2-2\hat{\xi} \\
\zeta & \sigma + \pi^*c_1(B_3)
\end{array}\end{equation}
We see, then, that ${\cal{C}}^{(3)}$ and ${\cal{C}}^{(2)}$ are in the classes
\begin{equation}{\cal{C}}^{(3)}=3\sigma + \pi^*(4c_1(B_3)-\hat{\xi})\quad {\cal{C}}^{(2)}=2\sigma + \pi^*(2c_1(B_3)+\hat{\xi})\end{equation}
We now discuss the "dual matter surfaces" for the various $\mathbf{10}$'s and $\mathbf{\overline{5}}$'s.

We start with $\mathbf{10}^{(3)}$, whose dual matter surface sits inside ${\cal{C}}^{(3)}$ and, by analogy with the $\mathbf{10}$ matter surface in the unfactored case, is given by
\begin{equation}(zv)=0\quad\tilde{d}_3=0\end{equation}
This means that
\begin{equation}\tilde{\Sigma}_{\mathbf{10}^{(3)}} = (\sigma + \pi^*S_2)\cdot \pi^*(c_1(B_3)-\hat{\xi})\end{equation}
The $\mathbf{10}^{(2)}$ dual matter surface is similar.  It is given by
\begin{equation}(zv)=0\quad \tilde{e}_2=0\end{equation}
and hence is in the class
\begin{equation}\tilde{\Sigma}_{\mathbf{10}^{(2)}} = (\sigma+\pi^*S_2)\cdot \pi^*\hat{\xi}\end{equation}

We now turn to the $\mathbf{\overline{5}}$ dual matter surfaces, starting with the one for $\mathbf{\overline{5}}^{(3)}$ which sits in ${\cal{C}}^{(3)}\cap \tau {\cal{C}}^{(3)}$.  The intersection ${\cal{C}}^{(3)}\cap \tau {\cal{C}}^{(3)}$ is given by
\begin{equation}\begin{split}0&=(zv)\left[\tilde{\alpha}\tilde{e}_0(zv)^2 + \tilde{d}_2\zeta^2\right] \\
0 &= \zeta\left[-\tilde{\alpha}\tilde{e}_1 (zv)^2 + \tilde{d}_3 \zeta^2\right]
\end{split}\end{equation}
The $\mathbf{\overline{5}}^{(3)}$ dual matter surface corresponds to the intersection of the two terms in $[\,]$'s less the $\zeta=\tilde{\alpha}=0$ and $\zeta=zv=0$ components.  As such, we have
\begin{equation}\begin{split}\tilde{\Sigma}_{\mathbf{\overline{5}}^{(3)}} &= {\cal{C}}^{(3)}\cdot {\cal{C}}^{(3)} - [zv]\cdot [\tilde{d}_3] - [\zeta]\cdot [\tilde{\alpha}\tilde{e}_0] - 2[\zeta]\cdot [\tilde{\alpha}] - 9[\zeta]\cdot [z] \\
&=  2\sigma\cdot \pi^*(3c_1(B_3)-2S_2)+8(\pi^*c_1(B_3))^2 + \pi^*\hat{\xi}\cdot \pi^*(S_2+\hat{\xi}) - \pi^*c_1(B_3)\cdot \pi^*(5S_2+3\hat{\xi})
\end{split}\end{equation}

Next, consider the $\mathbf{\overline{5}}^{(2)}$ dual matter surface, which sits inside ${\cal{C}}^{(2)}\cap \tau {\cal{C}}^{(2)}$.  The intersection ${\cal{C}}^{(2)}\cap \tau {\cal{C}}^{(2)}$ is described by
\begin{equation}\begin{split}
0 &= \tilde{e}_0 (zv)^2 + \tilde{e}_2 \zeta^2 \\
0 &= \tilde{e}_1 (zv)\zeta
\end{split}\end{equation}
This has several irreducible components.  The $\mathbf{\overline{5}}^{(2)}$ dual matter surface is obtained by removing the components $\zeta=\tilde{e}_0=0$, $\tilde{e}_2=(zv)=0$, and $\zeta=v=0$.  We therefore have
\begin{equation}\begin{split}\tilde{\Sigma}_{\mathbf{\overline{5}}^{(2)}} &= {\cal{C}}^{(2)}\cdot {\cal{C}}^{(2)} - [\zeta]\cdot [\tilde{e}_0] - [\tilde{e}_2]\cdot [zv] - 4[\zeta]\cdot [v] \\
&= 2\sigma\cdot \pi^*(c_1(B_3) - S_2 + \hat{\xi}) + (2c_1(B_3)+\hat{\xi})\cdot (c_1(B_3)-S_2+\hat{\xi})
\end{split}\end{equation}

Finally, we study the $\mathbf{\overline{5}}^{(3)(2)}$ dual matter surface, which sits inside ${\cal{C}}^{(2)}\cap \tau {\cal{C}}^{(3)}+{\cal{C}}^{(3)}\cap\tau{\cal{C}}^{(2)}$.  The intersection ${\cal{C}}^{(3)}\cap \tau {\cal{C}}^{(2)}$ is described by the equations
\begin{equation}\begin{split}0 &= \tilde{e}_0(zv)^2 - \tilde{e}_1(zv)\zeta + \tilde{e}_2\zeta^2 \\
0 &= \tilde{\alpha}(zv)^2\left[\tilde{e}_0 (zv) - \tilde{e}_1\zeta\right] + \tilde{d}_2 (zv)\zeta^2 + \tilde{d}_3 \zeta^3
\end{split}\end{equation}
Plugging the first into the second, we can rewrite this system as
\begin{equation}\begin{split}
0 &= \tilde{e}_0(zv)^2 - \tilde{e}_1(zv)\zeta + \tilde{e}_2\zeta^2 \\
0 &= -\tilde{\alpha}\tilde{e}_2\zeta^2(zv) + \tilde{d}_2 (zv)\zeta^2 + \tilde{d}_3\zeta^3
\end{split}\end{equation}
Here, there are components corresponding to $\zeta=\tilde{e}_0=0$ as well as the usual $\zeta=z=0$.  Removing these, we are left with that part of the $\mathbf{\overline{5}}^{(3)(2)}$ dual matter surface that sits inside ${\cal{C}}^{(3)}$.  We get a similar contribution from inside ${\cal{C}}^{(2)}$ so that the net $\mathbf{\overline{5}}^{(3)(2)}$ dual matter surface is
\begin{equation}\begin{split}\tilde{\Sigma}_{\mathbf{\overline{5}}^{(3)(2)}}&= 2\left[{\cal{C}}^{(3)}\cdot {\cal{C}}^{(2)} - 2[\zeta]\cdot[\tilde{e}_0] - 6[\zeta]\cdot[z]\right]\\
&= 2\left[\sigma\cdot \pi^*(4c_1(B_3)-2S_2-\hat{\xi}) + 4(\pi^*c_1(B_3))^2 - (\pi^*\hat{\xi})^2 - 2\pi^*c_1(B_3)\cdot \pi^*S_2\right]
\end{split}\end{equation}

\newpage

\bibliographystyle{JHEP}
\renewcommand{\refname}{Bibliography}
\addcontentsline{toc}{section}{Bibliography}

\providecommand{\href}[2]{#2}\begingroup\raggedright\endgroup


\end{document}